\documentclass[a4paper,11pt]{article}

\pdfoutput=1
\usepackage{subfigure}
\usepackage{booktabs}

\usepackage{lineno}
\usepackage[colorlinks=false]{hyperref}
\usepackage{bm}
\usepackage{graphicx}
\usepackage{amssymb}
\usepackage{amsmath}
\usepackage{tabularx}
\usepackage{natbib}
\usepackage{caption}
\usepackage{amsmath}
\usepackage{amsfonts}
\usepackage{amssymb}

\newcommand*{\abs}[1]{\left|{#1}\right|}
\newcommand*{\bracket}[1]{\left({#1}\right) } 

\modulolinenumbers[5]

\usepackage{authblk}

\usepackage[left=15mm,right=15mm,top=1.5cm,bottom=1.5cm,includeheadfoot]{geometry}
\setcitestyle{square,numbers}

\begin{document}

\title{Surrogate modeling of indoor down-link human exposure based on sparse polynomial chaos expansion}


\author[1,2]{Zicheng Liu}
\author[2]{Dominique Lesselier}
\author[3]{Bruno Sudret}
\author[1]{Joe Wiart}

\affil[1]{\scriptsize Chaire C2M, LTCI, T\'el\'ecom Paris, Paris 75013, France.}
\affil[2]{\scriptsize Laboratoire des Signaux et Syst\`emes, UMR8506 (CNRS-CentraleSupelec-Universit\'e Paris-Sud), Universit\'e Paris-Saclay, Gif-sur-Yvette cedex 91192, France}
\affil[3]{\scriptsize ETH Z\"urich, Chair of Risk, Safety and Uncertainty Quantification, Stefano-Franscini-Platz 5, Z\"urich 8093, Switzerland}

\maketitle

\abstract{
Human exposure induced by wireless communication systems increasingly draws the public attention. Here, an indoor down-link scenario is concerned and the exposure level is statistically analyzed. The electromagnetic field (EMF) emitted by a WiFi box is measured and electromagnetic dosimetry features are evaluated from the whole-body specific absorption rate as computed with a Finite-Difference Time-Domain (a.k.a. FDTD) code. Due to computational cost, a statistical analysis is performed based on a surrogate model, which is constructed by means of so-called sparse polynomial chaos expansion (PCE), where the inner cross validation (ICV) is used to select the optimal hyperparameters during the model construction and assess the model performance. However, the ICV error is optimized and the model assessment tends to be overly optimistic with small data sets. The method of cross-model validation is used and outer cross validation is carried out for the model assessment. The effects of the data preprocessing are investigated as well. Based on the surrogate model, the global sensitivity of the exposure to input parameters is analyzed from Sobol' indices.
\\

\noindent
\begin{tabular}{ll}
\textbf{Keywords:} &Specific absorption rate, surrogate model, polynomial chaos expansion,\\& least angle regression, orthogonal matching pursuit, cross-model validation,\\& double cross validation, Sobol' indices, global sensitivity analysis,\\& data preprocessing
\end{tabular}
}

\section{Introduction}
Human exposure \cite{wiart2016radio} to radio-frequency (RF) electromagnetic fields (EMF) increasingly draws public attention \cite{world2010research} due to the rapid development of wireless communications and the Internet-of-Things (IoT). The exposure by pervasive telecommunication devices and infrastructures needs to be evaluated. Research efforts have been carried out recently, in which the exposure level is evaluated by the computation of the specific absorption rate (SAR) \cite{wiart2016radio}, which quantifies the power absorbed by human tissues. The distribution of electric field inside the body is estimated by measurements with an equivalent phantom or by simulation with a numerical modeling approach, e.g., finite-difference in time-domain (FDTD) \cite{martinez2009fdtd} and finite-element methods (FEM) \cite{meyer2003human}.

According to the position of the human body under consideration, the investigation of electromagnetic dosimetry is often divided into indoor scenarios {\cite{findlay2010sar,findlay2012sar,ferreira2015sar,chiaramello2019children,chiaramello2019radio}} and outdoor ones {\cite{chobineh2019statistical,lacroux2008specific,aerts2018use,azzi2019surrogate}}, where near fields and far fields are respectively considered. Particular attention is given to indoor scenarios here, since people spend about $70\%$ of their time inside rooms \cite{zeghnoun2010description}. So, in comparison with far-field scenarios, the emitting source can be placed quite close to the human body and the exposure level may be high. Despite related researches mainly on deterministic investigation of several representative cases {\cite{findlay2012sar,findlay2010sar,varsier2014influence}}, a statistical analysis {\cite{wiart2016radio,chiaramello2017stochastic}} is necessary to estimate the human exposure with uncertain factors (e.g., frequency, distance) and exhibit the key impacting factors, especially in highly varying scenarios. 

Monte-Carlo simulations are often run for statistical analysis but may be intractable in our case due to the high computational/measurement cost to compute SAR. Recently, surrogate models based on the polynomial chaos expansions (PCE) have been constructed to approximate the complicated relation between the EMF exposure and the impacting factors \cite{liorni2016exposure,liorni2015study,pinto2017statistical}. Expanding the EMF exposure on orthogonal multivariate polynomials, the expansion coefficients are obtained as the ordinary least square (OLS) solution. To avoid the high variance of the constructed model and the overfitting problem, the cardinality of the basis should be constrained by setting a maximum value for the total degree of polynomials \cite{marelli2015uqlab}. However, the curse-of-dimensionality, i.e., the fact that the cardinality of the basis polynomially increases with the number of considered factors and the total degree, prevents broad applications of ordinary PCE-based surrogate models. Then, to reduce the basis size, approaches to construct so-called {\it sparse PCE model} \cite{blatman2011adaptive,blatman2010adaptive,marelli2015uqlab} have been proposed to only include the most influential polynomials in the PCE model. The influence of polynomials is usually measured by the correlation with the responses and the polynomials included in the sparse PCE model are selected by least angle regression (LARS) \cite{efron2004least} or orthogonal matching pursuit (OMP) \cite{pati1993orthogonal}. 

Two hyperparameters, the total degree of polynomials and the number of included polynomials in the final model, are involved in the construction of sparse PCE models \cite{marelli2015uqlab}. Their selection is performed via cross validation (CV). The hyperparameters corresponding to the smallest CV error are treated as optimal. This criterion of smallest CV error is also used to  assess the prediction performance of the final model. Therefore, CV plays the role for both model selection (hyperparameter selection) and model assessment. However, it has been recognized that if the final model is obtained via an extensive optimization by the CV and the performance is assessed by the same CV based on the same experimental design (ED), the model assessment tends to be optimistic \cite{anderssen2006reducing}. 

In the mentioned above study, over-optimistic model assessments appear with small training data sets, which include tens to hundreds of samples. So-called {\it cross-model validation} (CMV) \cite{anderssen2006reducing,baumann2014reliable} is used to overcome the bias of the assessment. Regarding the CV for the model selection as the first-layer validation, a second-layer CV is introduced to assess the constructed model on data independent of the training process. With the assessment of models constructed with LARS and OMP by CMV, the best model is chosen for predictions and inferences. Meanwhile, data preprocessing (e.g. representation of inputs in Cartesian or spherical coordinate system) significantly impacts the performance of the surrogate model. Such effects are investigated here.

{The present contribution focuses onto the statistical analysis of indoor down-link exposure based on the PCE model, whose assessment is based on the CMV technique. Therefore, much attention is given to CMV as key to reliable estimation of human exposure. That involves running examples with two analytical functions so as to demonstrate the effectiveness of CMV first, considering that this tool has indeed been proposed in investigations of Quantitative Structure-Activity Relationship (QSAR) [27] for variable selection, yet no literature is available, to the best of our knowledge, about its application to model assessment of PCE models themselves.}

This paper is organized as follows: the problem statement describing the down-link dosimetry case study is given in Section \ref{sec:ED}. Sparse PCE models based on LARS and OMP are considered in Section \ref{sec:PCE}. In Section \ref{sec:CMV}, the idea of CMV is introduced in detail and the assessment accuracy is illustrated by results in Section \ref{sec:results}, where the effect of data preprocessing is shown as well. 

\section{Experimental design (ED)}
\label{sec:ED}
\begin{figure}[!h]
	\centering
	\includegraphics[width=0.45\linewidth]{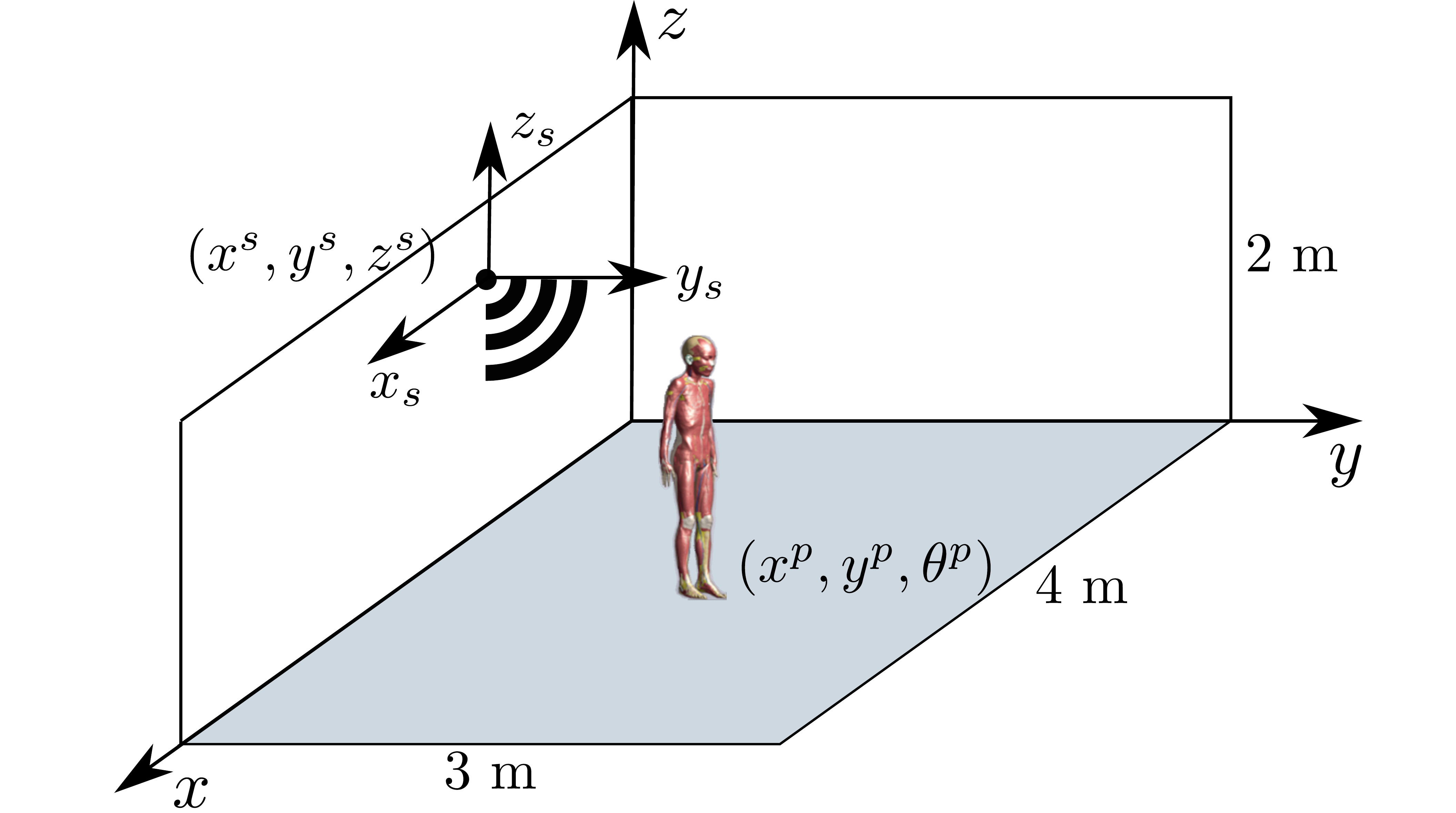}
	\caption{Uncertain parameters considered in the indoor down-link scenario.}
	\label{sketchFig}
\end{figure}

The scenario of interest is sketched in Fig.~\ref{sketchFig}, where a WiFi box working at $2,400$ MHz is moving along the wall of a $4\times3\times2 \, \text{m}^3$ room and a high-resolution female model, named as "Eartha" from the Virtual Family \cite{christ2009virtual}, is standing inside. The positions of the WiFi box and human model impact the electromagnetic exposure, which is quantified by the whole-body SAR, i.e., the averaged SAR over the whole body.

The averaged SAR over the body $\mathbb{V}$ is computed according to  
\begin{equation}
\text{SAR}=\frac{1}{V}\int_{\mathbb{V}}\frac{|\bm{E}(\bm{r})|^2\sigma(\bm{r})}{\rho(\bm{r})}d\bm{r},
\label{defSAR}
\end{equation}
where $V$ is the volume of $\mathbb{V}$, $\bm{E}(\bm{r})$ is the root-mean-square electric field, and $\sigma(\bm{r})$, $\rho(\bm{r})$ denote the conductivity and the tissue density at the sampling position $\bm{r}$. From Eq.\eqref{defSAR}, one sees that the uncertain factors influencing the SAR computation may come from the accuracy of measured/simulated electric fields, the physical properties of tissues, and the numerical computation of the integral in Eq.~\eqref{defSAR}. Here, only the uncertainty of the electric field due to the relative position of the source and the human model is considered. 

\begin{table}[!h]
	\centering
	\begin{tabular}{clcl}
		\toprule
		Source (notation) & Range & Human model (notation) & Range \\
		\midrule
		$x^s$ ($X_1$) & $[0.3, 3.7]$ m & $x^p$ ($X_4$) & $[0.05, 3.95]$ m\\
		$y^s$ ($X_2$) & $[0.3, 2.7]$ m & $y^p$ ($X_5$) & $[0.05, 2.95]$ m\\
		$z^s$ ($X_3$) & $[0.25, 2]$ m & $\theta^p$ ($X_6$) & $[0, 360)$ degree\\
		\bottomrule
	\end{tabular}
	\caption{Range of the uncertain variables.}
	\label{uncertainVari}
\end{table}

In the global Cartesian coordinate system, as shown in Fig.~\ref{sketchFig}, the position of the WiFi box and the one of the human model are denoted by $(x^s,y^s,z^s)$ and $(x^p,y^p,z^p)$, respectively. The human model is supposed to stand in the room, thus $z^p$ is considered a constant and not a uncertain variable. Its orientation may matter and is considered through the parameter $\theta^p$. Therefore, six uncertain variables are involved in the uncertainty analysis. Assuming the variables follow uniform distributions, the input space is defined by the ranges given in Table~\ref{uncertainVari}, where $0.25$ m is the minimum height of source, $0.3$ m and $0.05$ m are the minimum distance of the human model and the source to the wall, respectively. To efficiently characterize the response of the physical system with a limited number of samples, Latin hypercube sampling (LHS) \cite{wyss1998user} is used to determine the input samples while the corresponding whole-body SAR is computed according to the following procedures.

When the human body is far from the emitting source, the front of the incident wave is approximately planar and the source is often treated as planar waves \cite{hirata2007dominant,conil2011influence}. However, this approximation is not applicable in our case since the source may be quite close to the human body and the near field should be analyzed too. To characterize the near field scattered by the box, the measurement system StarLab by MVG$^{\text{\textregistered}}$ is used to measure the field. The field is expressed as a spherical wave expansion \cite{hald1988spherical}, with which an in-house FDTD code can be used to compute the field inside a high-resolution ($1$ mm$^3$) human model. However, the room size is too large to be directly considered in the FDTD code. The analyzed region is then reduced by computing a Huygens box enclosing the human model \cite{pinto2017statistical}. As a result, the computational burden becomes bearable with the utilization of GPU techniques and about 3 hours are required for each single deterministic computation. However, the computation cost is still too high to envisage statistical analysis by Monte-Carlo simulations. A surrogate model (or metamodel) is built in the next section to mimic the nonlinear relation between the whole-body SAR and six input parameters.

\section{Surrogate modeling based on polynomial chaos expansions}
\label{sec:PCE}

To facilitate the illustration, let us denote the six parameters gathered in Table~\ref{uncertainVari} by $\{{X}_i, i=1,2,\ldots,6\}$, the computed whole-body SAR by $Y$, and the resulting SAR $Y=\mathcal{M}(\bm{X})$. Assuming $E[Y^2]<+\infty$, the polynomial chaos expansion of the random SAR reads:
\begin{equation}
{Y}=\sum_{\bm{\alpha}\in\mathbb{N}^6}\beta_{\bm{\alpha}} {\Psi}_{\bm{\alpha}}(\bm{X}).
\label{completePCE}
\end{equation}
The construction of the basis follows the theory of generalized PCE (gPCE) \cite{xiu2002wiener,sudret2007uncertainty}. ${\Psi}_{\bm{\alpha}}$ is a multivariate polynomial, which is a tensor product of univariate polynomials, i.e.,
\begin{equation}
\psi_{\bm{\alpha}}(\bm{X}) = \pi_{{\alpha}_1}({X}_1) \times \ldots \times \pi_{{\alpha}_6}({X}_6),
\end{equation}
where $\bm{\alpha}=[\alpha_1,\ldots,\alpha_6]$, $\alpha_i$ as the degree of the univariate polynomial in variable $X_i$. The polynomial family $\pi_{{\alpha}_i}$ depends on the probability density function (PDF) of the random variable ${X}_i$. Since all input parameters have uniform distributions, Legendre polynomials are used to construct the basis polynomials in our case study. {The hyperparameter $\bm{\alpha}$ and the expansion coefficient $\beta_{\bm{\alpha}}$ are learned from the available input samples $\{\bm{x}^{(1)},\ldots,\bm{x}^{(N)}\}$ and corresponding output values $\{{y}^{(1)},\ldots,{y}^{(N)}\}$, $N$ being the size of the experimental design (ED).}

In practice, the infinite series of Eq.~\eqref{completePCE} shall be truncated and the most influential basis polynomials are to be included in the truncation. The commonly utilized approach is setting a maximum value to the total degree of basis polynomials \cite{blatman2010adaptive}, i.e., $\sum_{i=1}^{6}\alpha_i\le p$. Such a setting yields the so-called {\it full PCE} model. The value of $p$ decides for the flexibility of the PCE model. With a large $p$, the constructed model tends to be unbiased but with a high variance (especially when only a small set of data are available) and vice versa. An optimization is necessary to select the $p$ corresponding to the best model, given a particular set of runs. Here, ``best" is defined in terms of prediction accuracy, since the prediction ability is of interest. 

The assessment of prediction performance is often carried out by cross validation (CV), which includes the technique of leaving-many-out \cite{geisser1975predictive}, bootstrapping \cite{efron1992bootstrap} and $k$-fold CV \cite{arlot2010survey}. The latter is performed by dividing the whole set of data points into $k$ subsets with approximately the same size and by evaluating the surrogate model built from ($k-1$) subsets on the remaining $k$-th subset. With $k=N$, the $k$-fold CV becomes the so-called {\it leave-one-out CV} and the validation error is computed by
\begin{equation}
\epsilon^{\text{LOO}} = \frac{1}{N}\sum_{n=1}^{N}\bracket{\mathcal{M}(\bm{x}^{(n)})-\widehat{\mathcal{M}}^{-(n)}(\bm{x}^{(n)})}^2,
\end{equation}
where the superscript ``$(n)$" denotes the $n$-th sample, $n=1,2,\ldots,N$, and $\widehat{\mathcal{M}}^{-(n)}$ denotes the surrogate model trained by leaving the $n$-th sample out. This way, the value of $p$ corresponding to the minimum CV error is selected. 

Denote $\mathbb{A}^{\text{full}}$ as the set of selected $\bm{\alpha}$, $\bm{\Psi}_{\mathbb{A}^{\text{full}}}=[\Psi_{\bm{\alpha}},\bm{\alpha}\in\mathbb{A}^{\text{full}}]$, and $\bm{\beta}=[\beta_{\bm{\alpha}},,\bm{\alpha}\in\mathbb{A}^{\text{full}}]^T$, ``$T$" the transposition operator. The truncated PCE of Eq.~\eqref{completePCE} is written as
\begin{equation}
\mathcal{M}^{\text{full}}(\bm{X})=\sum_{\bm{\alpha}\in\mathbb{A}^{\text{full}}}\beta_{\bm{\alpha}} {\Psi}_{\bm{\alpha}}(\bm{X})=\bm{\Psi}_{\mathbb{A}^{\text{full}}}\bm{\beta}_{\mathbb{A}^{\text{full}}}.
\label{fullPCE}
\end{equation}
The ordinary-least-square (OLS) solution to the expansion coefficients is
\begin{equation}
\hat{\bm{\beta}}_{\mathbb{A}^{\text{full}}} = \bracket{\bm{\Psi}_{\mathbb{A}^{\text{full}}}^T\bm{\Psi}_{\mathbb{A}^{\text{full}}}}^{-1}\bm{\Psi}_{\mathbb{A}^{\text{full}}}^T\bm{y},
\label{OLS}
\end{equation}
where $\bm{y}$ is the vector gathering the SAR values computed for each input vector $\bm{x}^{(i)}$, $i=1,\ldots,N$. With $\bm{\Psi}_{\mathbb{A}^{\text{full}}}$ and $\hat{\bm{\beta}}_{\mathbb{A}^{\text{full}}}$, the full PCE model is constructed with Eq.~\eqref{fullPCE}. 
However, the cardinality of $\mathbb{A}^{\text{full}}$ equals ${p+6\choose p}$, which increases polynomially with $p$. As a result, the construction process of full PCE models suffers from the curse-of-dimensionality \cite{friedman2001elements}, which means that the required ED size shall become large to avoid the high variance (or the overfitting problem) of the constructed model. Then, so-called {\it sparse polynomial chaos expansions} \cite{blatman2010adaptive,blatman2011adaptive} have been proposed to reduce the basis.

\begin{table}[!h]
	\caption{Sparse PCE model based on orthogonal matching pursuit.}
	\label{alg:OMP}
	\begin{tabular}{r l}
		\toprule
		&For $p=1,2,\ldots,p_{\text{max}}$,\\[4pt]
		\hspace{-5pt}
		1.& Initialization: residual $\bm{R}_0=\bm{y}$, active set $\mathbb{A}_0^a=\varnothing$, candidate set $\mathbb{A}_0^c=\mathbb{A}^{\text{full}}_p$.\\[4pt]
		2.& For $j=1,2,\ldots,P_{\text{max}}=\min\{N-1,\text{card}(\mathbb{A}^{\text{full}}_p)\}$, \\
		& \hspace{2pt} 1) Find the basis element most correlated with $\bm{R}_{j-1}$:  ${\bm{\alpha}_j}=\arg \max_{\bm{\alpha}\in\mathbb{A}_{j-1}^c} \abs{\bm{R}_{j-1}^T\bm{\psi}_{\bm{\alpha}}}$.\\
		& \hspace{2pt} 2) Update $\mathbb{A}_j^a=\mathbb{A}_{j-1}^a\cup\bm{\alpha}_j$ and $\mathbb{A}_j^c=\mathbb{A}_{j-1}^c\setminus\bm{\alpha}_j$.\\
		& \hspace{2pt} 3) With $\bm{\psi}_{\mathbb{A}_j^a}$, compute $\bm{\beta}_j$ as the OLS solution.\\
		& \hspace{2pt} 4) Update residual $\bm{R}_j=\bm{y}-\bm{\psi}_{\mathbb{A}_j^a}^T\bm{\beta}_j$.\\
		& \small End\\[4pt]
		3.& Based on $\bm{\psi}_{\mathbb{A}_j^a}$ and $\bm{\beta}_j$, compute $\epsilon^{\text{LOO}}_{j}$, $j=1,2,\ldots,P_{\text{max}}$. \\[4pt]
		4.& $\epsilon^{\text{min}}_p=\min\{\epsilon^{\text{LOO}}_{j}\}$, ${P}=\arg\min_j \epsilon^{\text{LOO}}_{j}$ and the PCE model corresponding to $\bm{\psi}_{\mathbb{A}_{P}^a}$ is selected.\\[4pt]
		5.& For $p>2$, if $\epsilon^{\text{min}}_p>\epsilon^{\text{min}}_{p-1}>\epsilon^{\text{min}}_{p-2}$, stop.\\[4pt]
		&\hspace{-5pt}
		End\\
		\bottomrule
	\end{tabular}\\[-10pt]
\end{table}
Measuring the influence of multivariate polynomials by the correlation with the sampled output data, only the most correlated polynomials are included in the sparse PCE model. Matching pursuit algorithms, e.g., orthogonal matching pursuit (OMP) \cite{pati1993orthogonal} and least angle regression (LARS) \cite{efron2004least}, have been applied to rank and select the basis polynomials. The procedures to construct sparse PCE models based on OMP are summarized in Table~\ref{alg:OMP} \cite{liu2018surrogate}, where the number of included polynomials is smaller than $N-1$ to avoid overfitting modeling and the optimal number is decided by cross validation. The maximum value of $p$ is denoted as $p_{\text{max}}$ and the optimal value is decided by an early-stop criterion, i.e., when the CV error increases for two subsequent values of $p$, the algorithm will stop to avoid the overfitting problem. The model construction with LARS follows a similar procedure and is shown less greedy, since the data residual $\bm{R}$ evolves with equiangular directions of the basis instead of the basis itself \cite{liu2018surrogate,efron2004least}.  

\section{Cross-model validation}
\label{sec:CMV}

{As mentioned in Section \ref{sec:PCE}, a number of candidate surrogate models is constructed with the training data set by tuning the hyperparameters of the training process, i.e., the number of included polynomials in the final model $P$ and the total degree of multivariate polynomials $p$. With the model selection by cross validation, the surrogate model with the smallest validation error ($\epsilon^{\text{LOO}}$ here) is chosen as the final model.
	
The best model in terms of a given training process has the smallest validation error in ideal cases. However, the possibility for a candidate PCE model getting underestimated assessments may lead to a validation error even smaller than the error with the ``true" model. Consequently, this candidate model would be selected despite that it is suboptimal. Moreover, if this smallest validation error is used as the metric of model assessment, which is the case for some applications of sparse PCE models \cite{pinto2017statistical,kersaudy2015new}, the finally selected surrogate model would be over-optimistically assessed. This phenomenon is the so-called {\it model-selection bias} and can be detected by comparing the above validation error for model selection and the testing error which is obtained from the prediction on a data set independent of the training process. The bias is detected if the validation error is much smaller than the testing error.}

{The phenomenon of model-selection bias happens in the research of Quantitative Structure-Activity Relationship (QSAR) \cite{roy2015primer}, where a suitable subset of a high number of molecular descriptors is selected to optimize the model representation. The technique of cross-model validation (CMV) \cite{anderssen2006reducing} (or double cross validation \cite{baumann2014reliable}) was proposed with the knowledge that the data for model assessment should not be involved in the model construction or selection.

CMV introduces a second-layer cross validation, which is so-called {\it outer CV}, while the CV for model selection is {\it inner CV}. In the outer CV, the experimental design is randomly divided into two disjoint data sets, one for model assessment, the other for model construction and selection. This way, the data set for model assessment is independent of the training process. Thus, the outer CV error is an unbiased metric for model assessment. 

Approaches including the holdout method \cite{arlot2010survey}, leave-many-out, and $k$-fold CV are commonly utilized to split ED. The leave-one-out CV is applied in the outer CV considering the following aspects. First, most of available data are used for training in LOOCV to improve the surrogate-modeling performance, especially considering that cases with a quite small ED are of concern here. In addition, the outer CV actually assesses surrogate models constructed from subsets of the ED. Strictly speaking, the final model trained with the whole set of available data is not validated and the assessment by outer CV tends to be pessimistic due to less training data. Since only a single sample is left out from the training process, outer LOOCV can well approximate the performance of the final model.         

\begin{figure}[!h]
	\centering
	\includegraphics[width=0.6\linewidth]{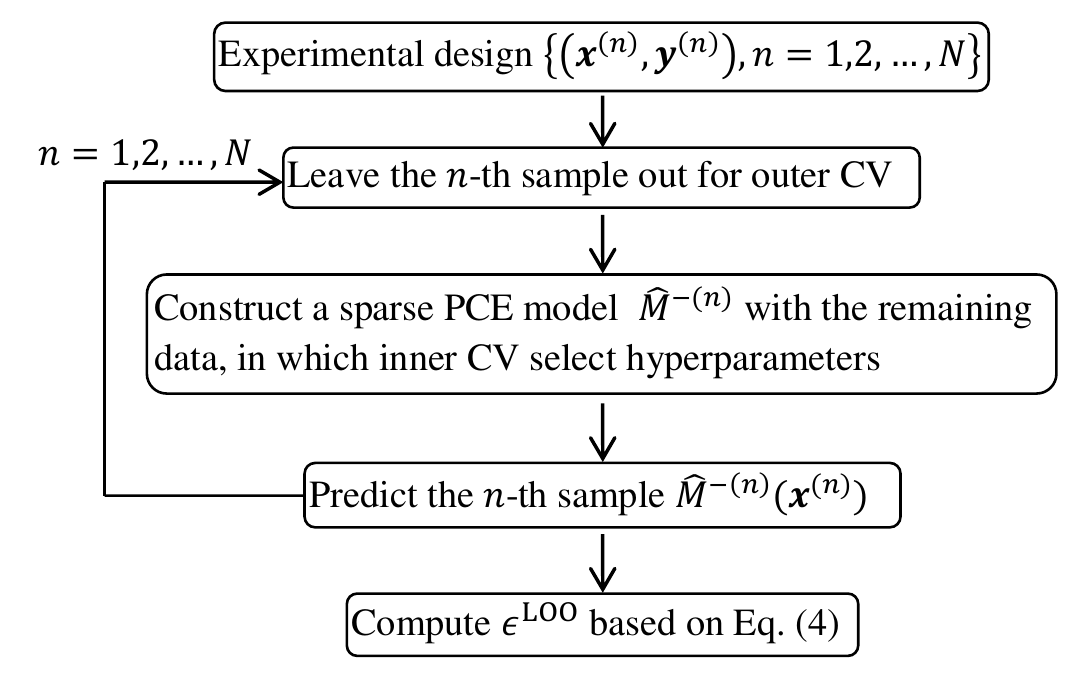}
	\caption{Flow chart of outer leave-one-out cross validation.}
	\label{flowChart}
\end{figure}

Fig.~\ref{flowChart} gives the flow chart performing the outer CV by leaving a single sample out. Remark that, although each sample is sequentially left out for outer CV, actually no sample is totally independent of the training process. However, the mode-selection bias is absent since the sample for outer CV is not involved in the model construction and selection in every single data division.   

Due to the introduction of outer LOOCV, $N$ surrogate models are built to compute the outer CV error and about $N-1$ times more computations are required. Since cases with small ED are of concern, the construction of a single surrogate model is usually efficient and the running of outer CV is tractable. Otherwise, after the division of ED, since the model construction with different training data sets is independent of each other, parallel computational techniques (e.g., distributed computation) can be used.}    

\section{Results}
\label{sec:results}

The sparse {and full} PCE models are constructed with the open-source software UQLab V1.2.0 \cite{marelli2015uqlab}, in which LARS or OMP is used to {build sparse PCE models}. Default configurations of UQLab for PCE models are adopted, except that the value of $p_{\max}$ in Algorithm \ref{alg:OMP} is set at $10$. This section shows the model assessment by inner CV, CMV and independent testing. The improved performance based on pre-processing the data and the results of sensitivity analysis are shown in the second subsection. { Let us emphasize that for the following examples, constructing a sparse PCE model or a full PCE model with a standard personal computer takes less than 10 seconds, so only the prediction accuracy is of concern.}

\subsection{Validation}
The accuracy of model assessments by inner CV and outer CV (or CMV) is checked by the validation on independent testing data, the size of which should be large enough to ensure the testing accuracy.  For the concerned scenario, since the size of the available data is limited, a subset (e.g., half) of the data is arbitrarily drawn for independent testing. The over-optimistic assessment by inner CV is independent of the considered scenarios. Two analytical functions, namely the Ishigami function and borehole function, are analyzed. In these two cases, a large set of testing data is easily achievable.   

According to the above illustration, one has a three-layer validation including inner CV, outer CV and independent testing. Inner CV is commonly used to do both hyperparameter selection and model assessment. In the framework of cross-model validation, the model assessment is performed by outer CV instead. The assessment accuracy is examined by the independent testing. 

The validation error corresponding with the three-layer validation is denoted by $\epsilon_{\text{type}}$, where the subscript is ``ICV" fo inner CV, ``OCV" for outer CV or ``test" for testing. Here, $\epsilon_{\text{ICV}}$ is withdrawn from UQLab, in which the LOOCV is performed and the validation error is corrected \cite{chapelle2002model} by
\begin{equation}
\epsilon_{\text{ICV}} = \epsilon^{\text{LOO}}\bracket{1-\frac{\text{card}(\mathbb{A})}{N}}^{-1}\bracket{1+\text{tr}((\bm{\Psi}^T\bm{\Psi})^{-1})},
\end{equation}
where ``card" and ``tr" means the cardinality of a set and the trace of a matrix. Based on the validation error, the determination coefficient is computed as
\begin{equation}
Q^2_{\text{type}} = 1-\frac{\epsilon_{\text{type}}}{\text{var}(\bm{y})},
\end{equation} 
which indicates that with a better surrogate model, the associated $Q^2$ is closer to $1$. If the inner and outer CV can properly assess the PCE model, we would expect that $Q^2_{\text{ICV}}\approx Q^2_{\text{test}}$ and $Q^2_{\text{OCV}}\approx Q^2_{\text{test}}$.

\subsubsection{Ishigami function}
The Ishigami function \cite{ishigami1990importance}, which is widely used for benchmarking in uncertainty quantification and sensitivity analysis, is defined by
\begin{equation}
Y = \sin X_1 + a \sin^2 X_2 + bX_3^4\sin X_1,
\label{defIshigami}
\end{equation} 
where $a=7$, $b=0.1$ here. All random variables $X_i$, $i = 1,2,3$, are independent and uniformly distributed over $[-\pi, \pi]$. Legendre polynomials are used to compose the basis according to the principle of the generalized PCE. 

\begin{figure}[!ht]
	\centering
	\subfigure[LARS]{\includegraphics[width = 0.5\linewidth]{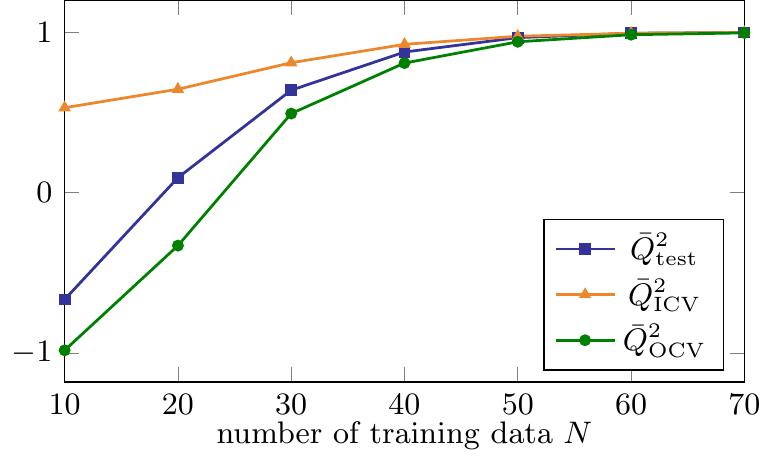}\label{Ishigami_CMV_LARS_Ntrain_LOO_LineGraph}}~
	\subfigure[OMP]{\includegraphics[width = 0.5\linewidth]{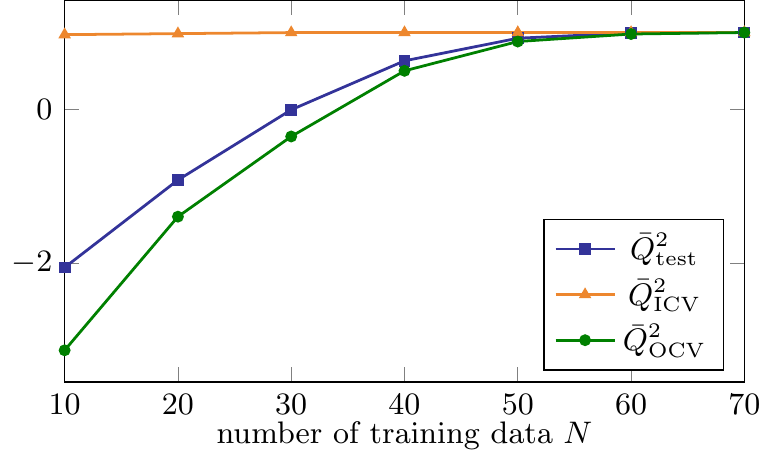}\label{Ishigami_CMV_OMP_Ntrain_LOO_LineGraph}}
	\caption{Ishigami function - mean of $Q^2$ over $100$ replications versus the increasing number of training data.}
	\label{Ishigami_CMV_Ntrain_LOO_LineGraph}
\end{figure}

Samples in the input space are obtained by Latin hypercube sampling (LHS). The testing data are composed of $10^4$ samples, while the number of training data $N$ increases from $10$ to $70$ and $100$ replications are performed for each $N$. The obtained validation results are shown in Fig.~\ref{Ishigami_CMV_Ntrain_LOO_LineGraph}, where $\bar{Q}^2$ denotes the averaged quantity over the replications. As observed, no matter with LARS or OMP, the inner CV tends to optimistically assess the constructed model and over-optimistic assessments happen when $N\le 40$, {where the variance of the surrogate modeling is high and the possibility for a candidate model having smaller (underestimated) validation error than the true model is large}. In contrast, the outer CV is likely to have slightly pessimistic assessments and the bias extent decreases as the performance of the surrogate model improves.   

\subsubsection{Borehole function}
\label{subsec:Borehole}
\begin{table*}[!ht]
	\centering
	\begin{tabular}{llll}
		\toprule
		Name & {Distribution}&{Bounds}& Description\\
		\midrule
		$r_w$ ($\text{m}$) & $\mathcal{N}(0.10, 0.0161812)$ & $[0.05, 0.15]$ & radius of borehole\\
		$r$ ($\text{m}$) & Lognormal(7.71, 1.0056) & $[100, 50000]$ & radius of influence\\
		$T_u$ ($\text{m}^2$/yr) & Uniform & $[63070, 115600]$ & transmissivity of upper aquifer\\
		$H_u$ ($\text{m}$) & Uniform & $[990, 1110]$ & potentiometric head of upper aquifer\\
		$T_l$ ($\text{m}^2$/yr) & Uniform & $[63.1, 116]$ & transmissivity of lower aquifer\\
		$H_l$ ($\text{m}$) & Uniform & $[700, 820]$ & potentiometric head of lower aquifer\\
		$L$ ($\text{m}$) & Uniform & $[1120, 1680]$ & length of borehole\\
		$K_w$ ($\text{m}$/yr) & Uniform & $[9855, 12045]$ & hydraulic conductivity of borehole\\
		\bottomrule
	\end{tabular}
	\caption{Borehole function - description and distribution of input variables.}\label{Borehole_pdf}
\end{table*}

The Borehole funtion \cite{harper1983sensitivity} models the water flow through a borehole and is defined as
\begin{equation}
Y = \frac{2\pi T_u(H_u-H_l)}{\ln (r/r_w)\bracket{1+{T_u}/{T_l}}+{2LT_u}/r_w^2K_w}
\end{equation}
The description and distribution of the involved eight independent variables are given in Table~\ref{Borehole_pdf}. While Legendre polynomials are used for the variables with uniform distributions, Hermite polynomials are used for the other variables after the isoprobabilistic transformation into standard normal variables \cite{lebrun2009generalization}. 

\begin{figure}[!ht]
	\centering
	\subfigure[LARS]{\includegraphics[width = 0.5\linewidth]{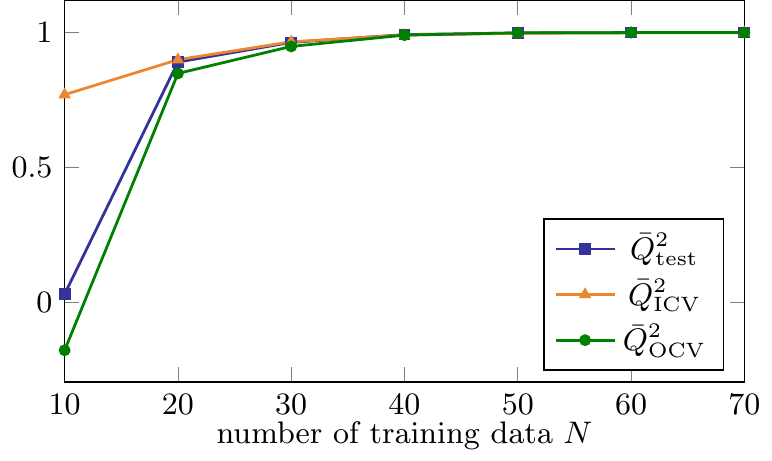}\label{Borehole_CMV_LARS_Ntrain_LOO_LineGraph}}~
	\subfigure[OMP]{\includegraphics[width = 0.5\linewidth]{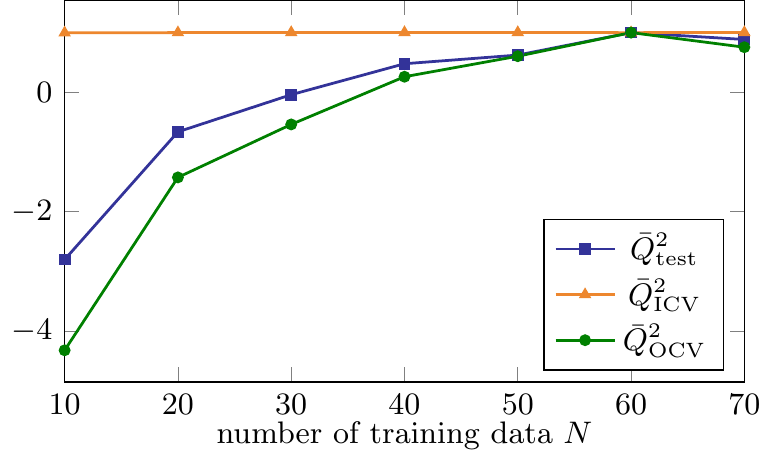}\label{Borehole_CMV_OMP_Ntrain_LOO_LineGraph}}
	\caption{Borehole function - mean of $Q^2$ over $100$ replications versus the increasing number of training data.}
	\label{Borehole_CMV_Ntrain_LOO_LineGraph}
\end{figure}

Sampling the input space by LHS method and taking $10^4$ data for independent testings, the determination coefficients computed by the three-layer validation are given in Fig.~\ref{Borehole_CMV_Ntrain_LOO_LineGraph}, where the optimistic assessments by ICV and pessimistic ones by OCV are observed. The over-optimistic assessment by ICV occurs when the constructed surrogate model has a poor performance, which happens when $N=10$ with LARS and $N=10,\ldots,50$ with OMP. In contrast, the outer CV slightly underestimates poor surrogate models but can properly assess the ones with high $Q^2$. Thus, one may conclude that the outer CV fits the practical requirements for the model assessment, since poor surrogate models can be prevented from the final application.

\subsubsection{Specific absorption rate}
\label{subsubsec:SAR}

\begin{figure}[!ht]
	\centering
	\subfigure[LARS, $R^2=0.9460$]{\includegraphics[width = 0.33\linewidth]{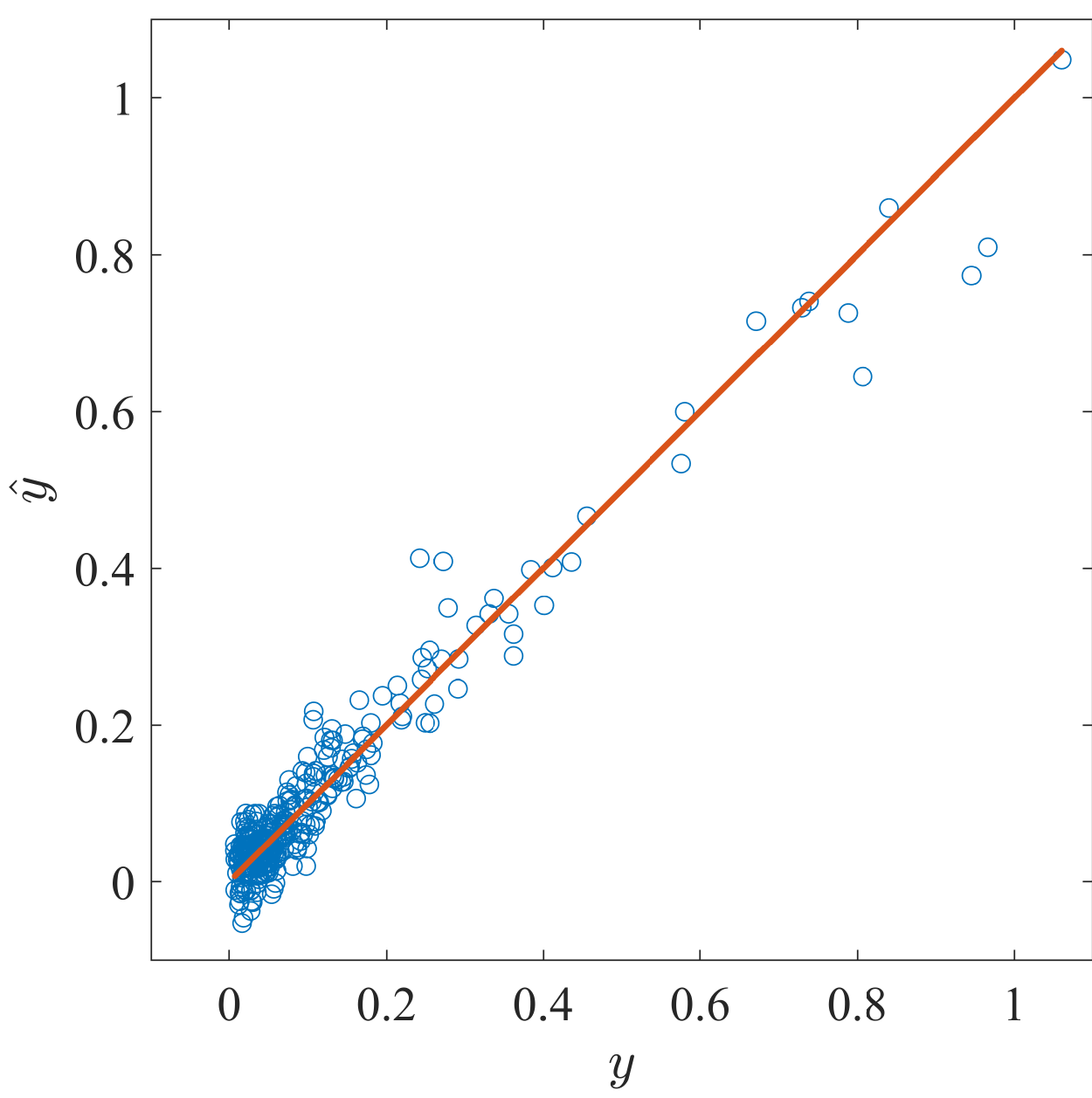}\label{Prediction_LARS_fullData}}~
	\subfigure[OMP, $R^2=1.0000$]{\includegraphics[width = 0.33\linewidth]{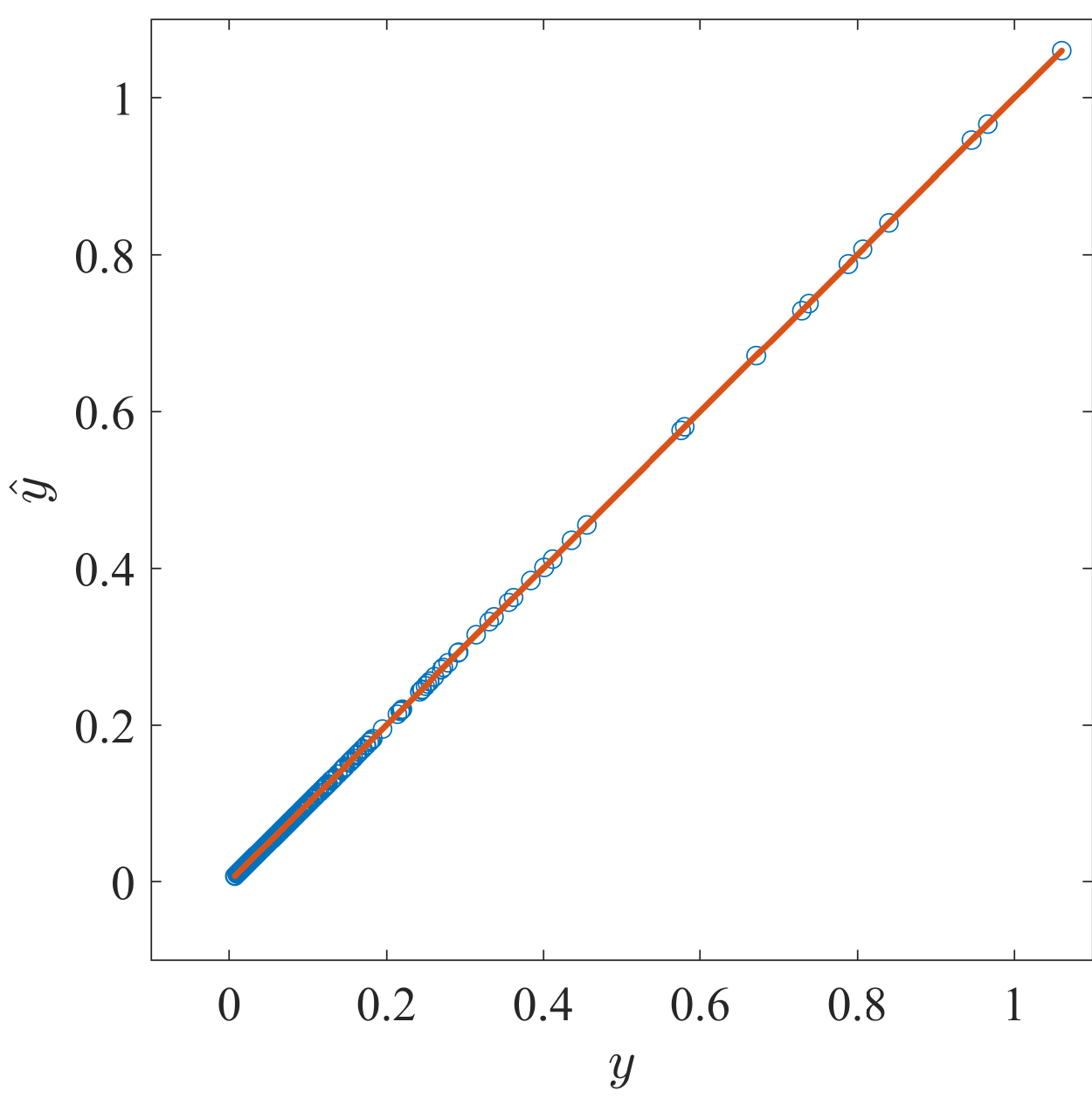}\label{Prediction_OMP_fullData}}~
	\subfigure[Full model, $R^2=0.4718$]{\includegraphics[width = 0.33\linewidth]{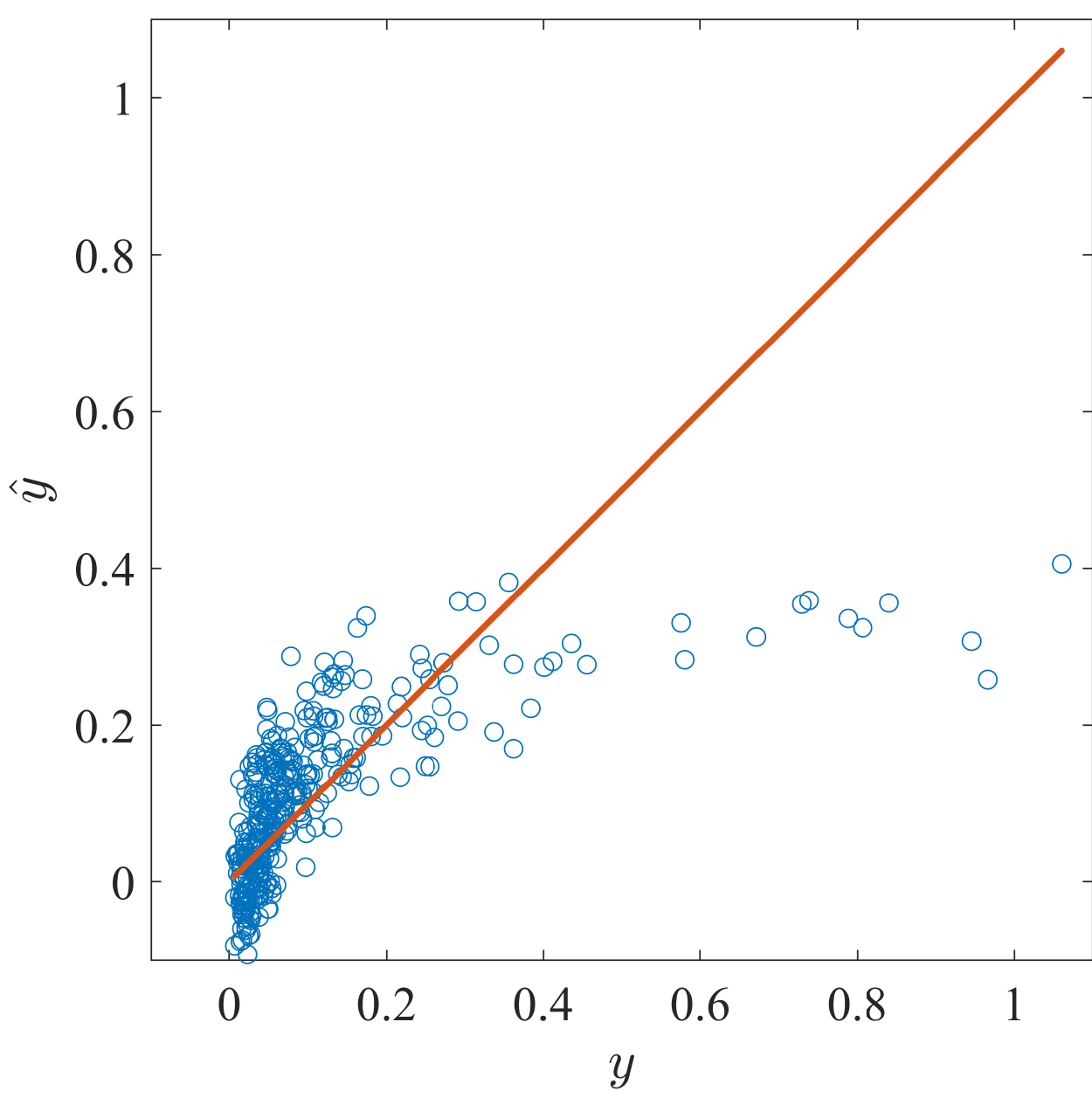}\label{Prediction_OLS_fullData}}
	\caption{Specific absorption rate - prediction of training data with the constructed PCE model based on the whole set of data.}
	\label{Prediction_fullData}
\end{figure}
Constructing the sparse PCE model with the whole set of data, which includes $350$ samples, the prediction of the training data is shown in {Fig.~\ref{Prediction_fullData}\subref{Prediction_LARS_fullData} and \subref{Prediction_OMP_fullData}}. As seen, the surrogate models based on LARS and OMP seem to have excellent prediction performance. Especially, the model based on OMP can accurately predict the training samples with {$R^2=1.0000$, which is defined by
	\begin{equation}
	R^2 = 1-\frac{\epsilon_{\text{train}}}{\text{var}(\bm{y})}, \, \epsilon_{\text{train}} = \frac{1}{N}\sum_{n=1}^{N}\bracket{\mathcal{M}(\bm{x}^{(n)})-\widehat{\mathcal{M}}(\bm{x}^{(n)})}^2.
	\end{equation}
	Thus, the prediction of the training dataset is accurate with OMP. In contrast, the full model, as shown in Fig.~\ref{Prediction_fullData}\subref{Prediction_OLS_fullData}, performs badly with $R^2=0.4718$. Remark that the full model is constructed with UQLab, where the maximum degree of polynomial is decided by the same early-stop criterion mention above. For this case, the underfitting full model is obtained due to the problem of insufficient modeling flexibility (or small $p$).}

\begin{figure}[!ht]
	\centering
	\subfigure[LARS, $78$ nonzero coefficients ]{\includegraphics[width = 0.37\linewidth]{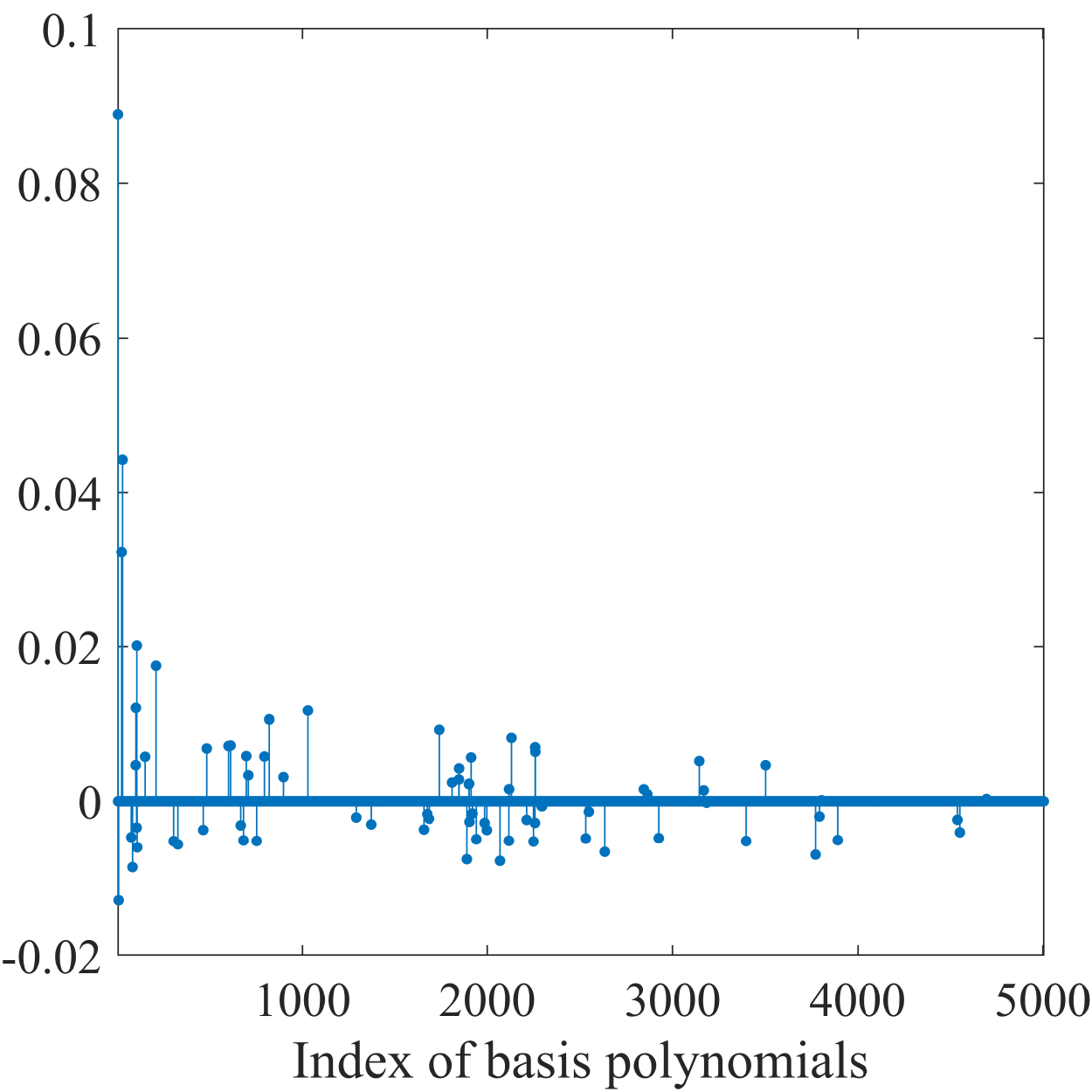}\label{Coeff_LARS_fullData}}~
	\subfigure[OMP, $308$ nonzero coefficients]{\includegraphics[width = 0.37\linewidth]{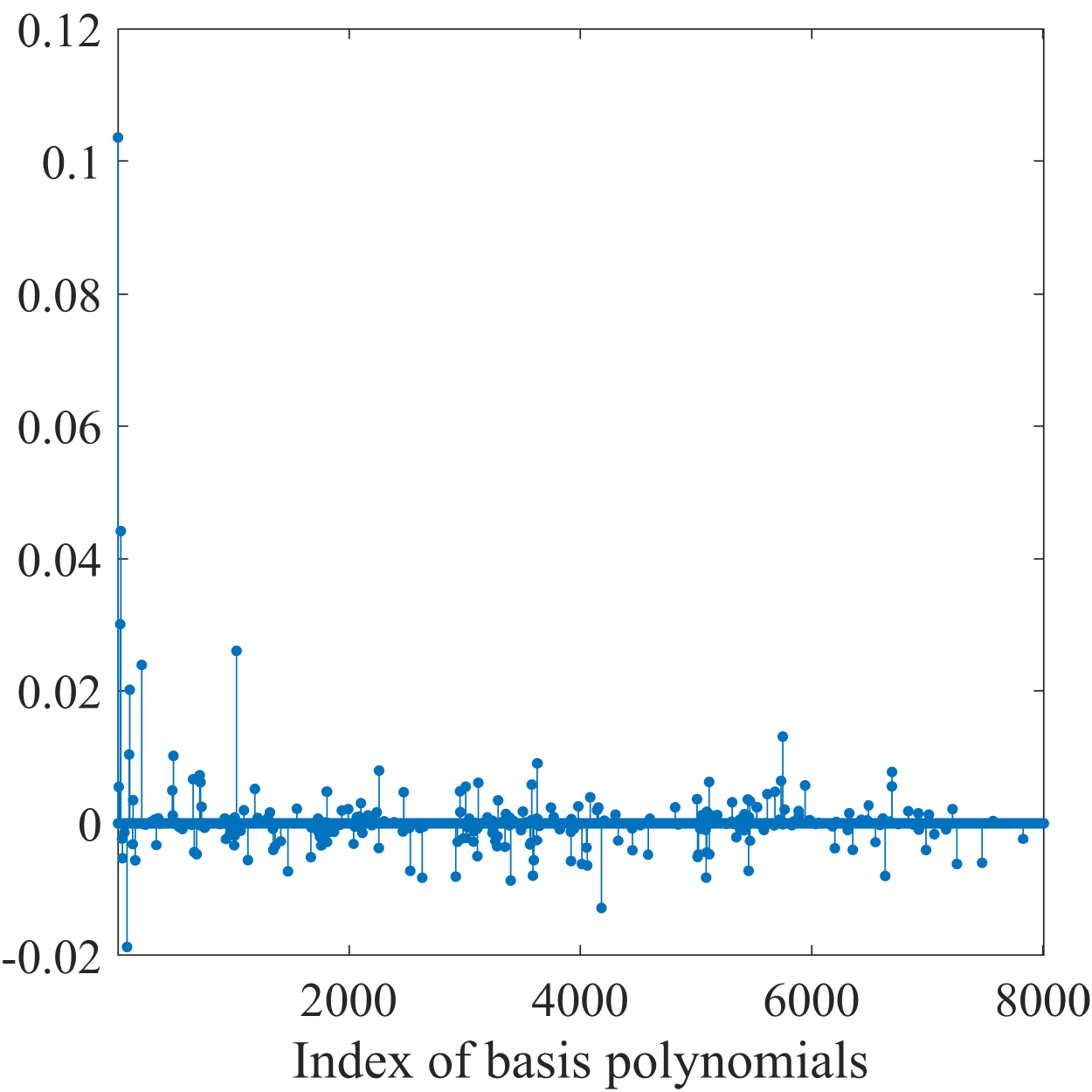}\label{Coeff_OMP_fullData}}
	\caption{Specific absorption rate - expansion coefficients of the constructed sparse PCE models.}
	\label{Coeff_FullSet}
\end{figure}
{Expansion coefficients of the sparse PCE models are plotted in Fig.~\ref{Coeff_FullSet} to show the sparsity on the selection of polynomials. With LARS, the optimal degree $p$ is chosen as $9$ by the early stop procedure. $5005$ basis polynomials are generated from the full model and $78$ ones are considered influential. With OMP, $p$ reaches $10$ and $308$ out of $8008$ basis polynomials are selected by the construction process.} 

In some applications of sparse PCE models, since the inner validation error is so small, the constructed model is to be used to predict unknown inputs and perform inferences. However, since the data for ICV already participated in the training process, the performance assessment should be checked further in case of over-optimistic assessments. {Remark that the performance of the full model is not checked further in the following part since the poor prediction on the training dataset in Fig.~\ref{Prediction_fullData}\subref{Prediction_OLS_fullData} already reveals the underfitting problem.} 

\begin{figure}[!ht]
	\centering
	\subfigure[LARS, $\bar{Q}^2_{\text{ICV}}=0.7490$ and $\bar{Q}^2_{\text{test}}=0.4203$]{\includegraphics[width = 0.35\linewidth]{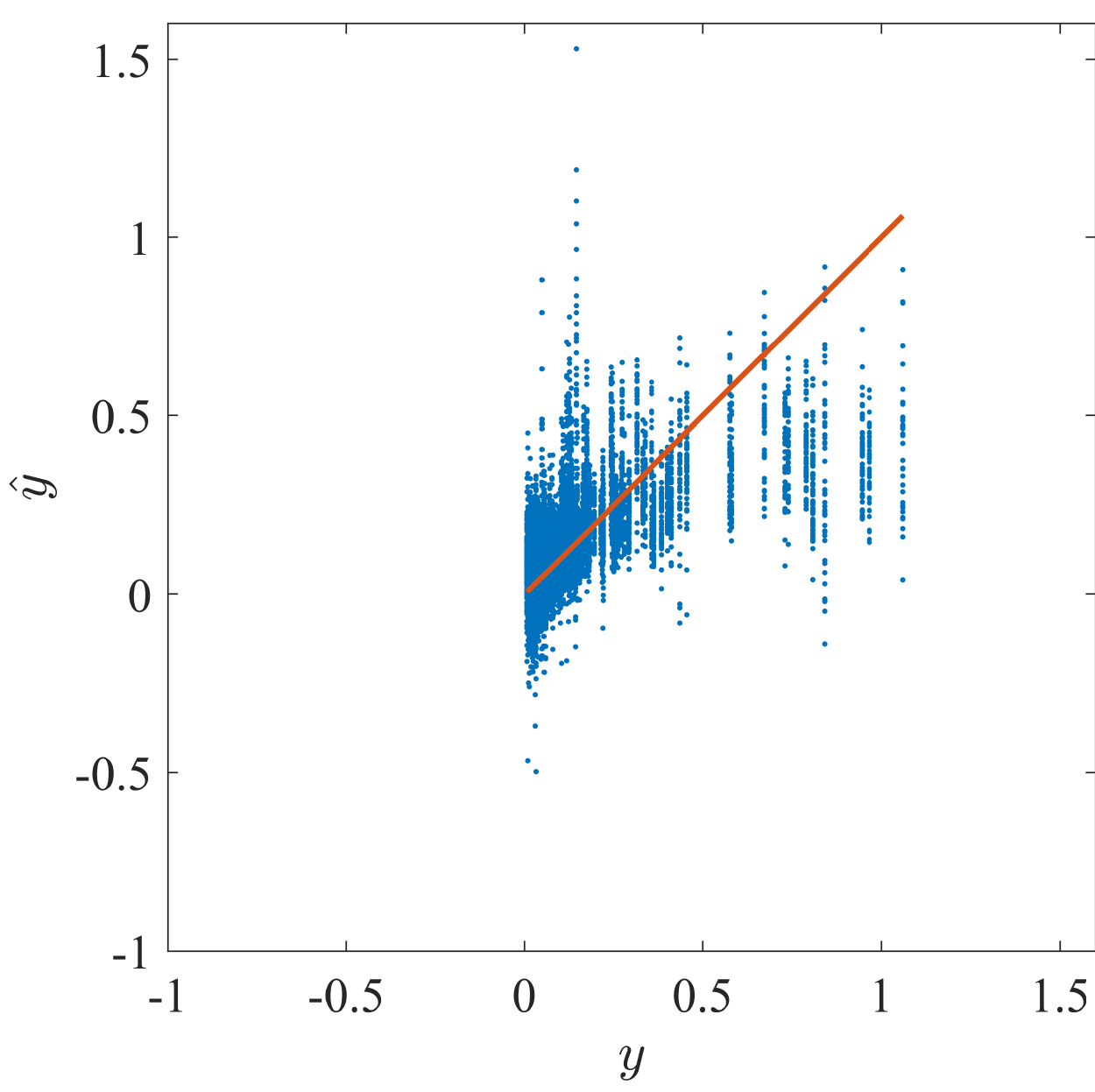}\label{Prediction_CMV_LARS_halfData}}~
	\subfigure[OMP, $\bar{Q}^2_{\text{ICV}}=1.0000$ and $\bar{Q}^2_{\text{test}}=-0.1262$]{\includegraphics[width = 0.35\linewidth]{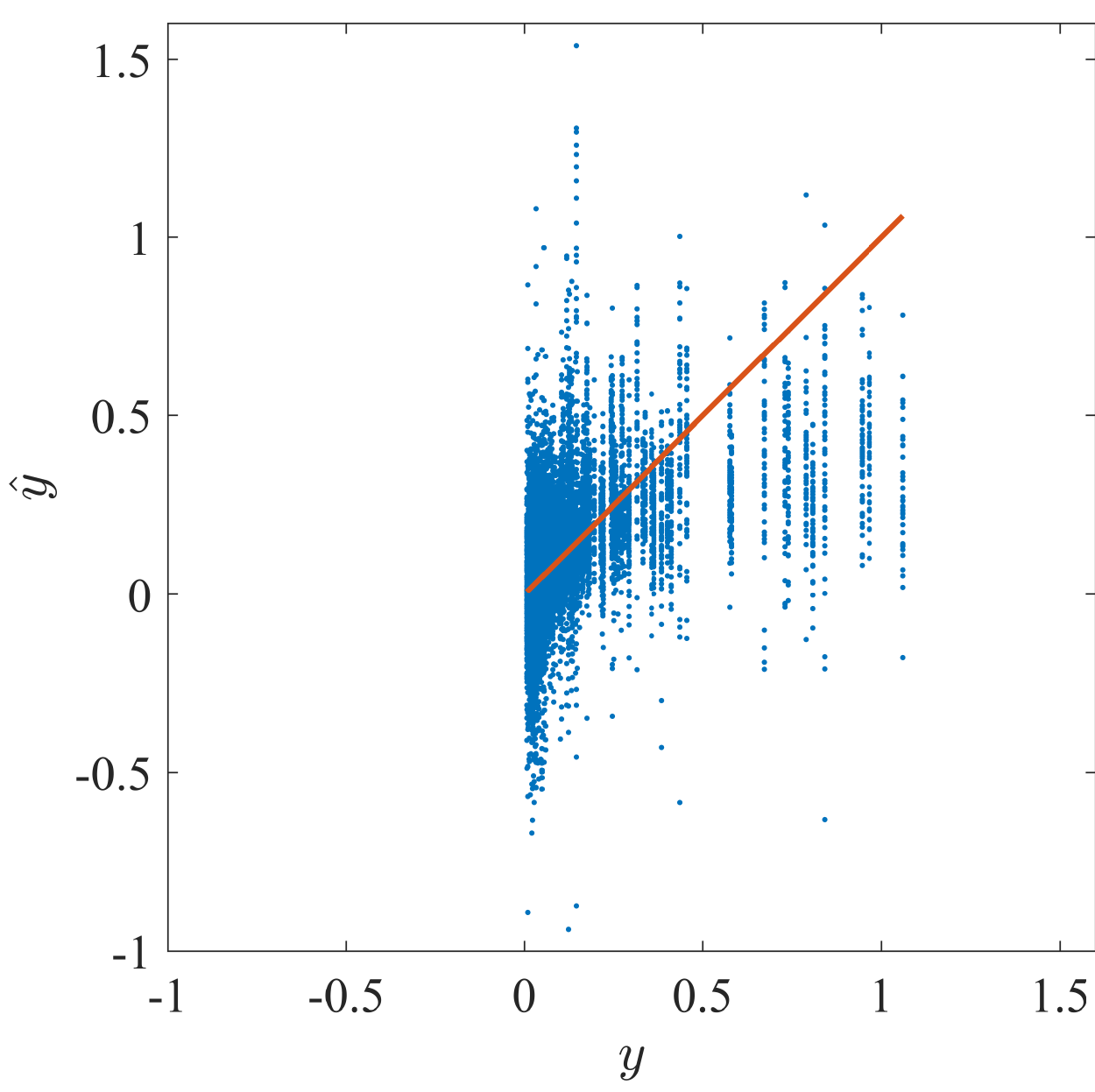}\label{Prediction_CMV_LARS_halfData}}
	\caption{Specific absorption rate - prediction of independent testing data with the sparse PCE model based on an half set of data}
	\label{Prediction_CMV_halfData}
\end{figure}

Randomly choosing half of the data set for independent testing, sparse PCE models are constructed with the remaining data. Repeating this process $100$ times, the prediction of $17,500$ resampled samples is given in Fig.~\ref{Prediction_CMV_halfData}. Although both approaches perform poorly, surrogate models based on LARS seem better than those based on OMP in the prediction of small whole-body SARs. The averaged $Q^2_{\text{test}}$ supports the same conclusion. However, since $\bar{Q}^2_{\text{ICV}}=1$, the OMP-based PCE model would be considered as an accurate representation of the original FDTD model. One may conclude that the assessment by inner CV can be quite misleading since it can be over-optimistic. {Moreover, the prediction performance of training data is much better than the testing data, i.e., $R^2\gg \bar{Q}^2_{\text{test}}$. Such phenomenon reveals overfitting.}

\begin{figure}[!ht]
	\centering
	\subfigure[LARS]{\includegraphics[width = 0.5\linewidth]{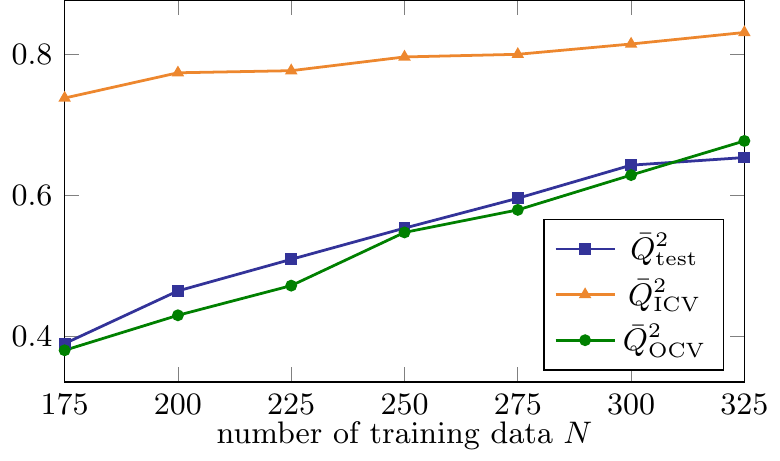}\label{CMV_LOO_LARS_maxOrder10_6Input}}~
	\subfigure[OMP]{\includegraphics[width = 0.5\linewidth]{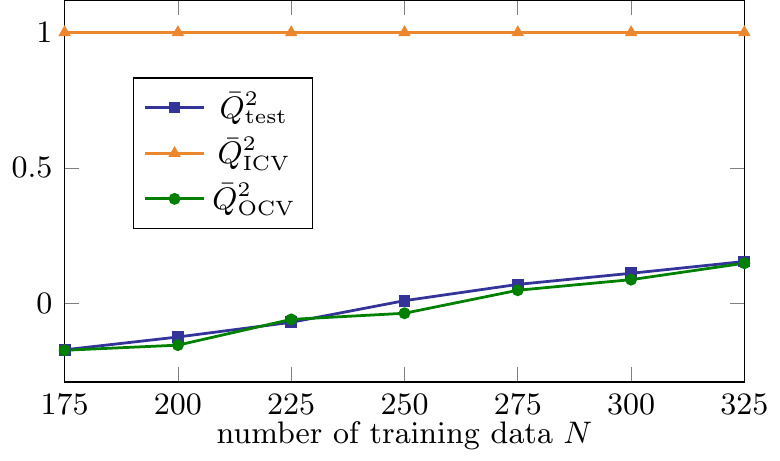}\label{CMV_LOO_OMP_maxOrder10_6Input}}
	\caption{Specific absorption rate - mean of $Q^2$ over $100$ replications versus the increasing number of training data.}
	\label{Prediction_CMV_LOO}
\end{figure}
Since the size of the available data set is usually small, the number of data for model assessment is preferred to be small so that most of the data is used in the training process. Therefore, the outer CV in cross-model validation is operated by leave-one-out CV. To check the accuracy of outer CV, a portion of data is used for testing. Repeating the training and testing process $100$ times, the average determination coefficients are plotted in Fig.~\ref{Prediction_CMV_LOO} as a function of the size of training data set, $N$, which increases from $175$ to $325$ while the size of testing data set equals $350-N$. As observed from the curve of $\bar{Q}_{\text{test}}^2$, the performance of the constructed surrogate model is improved when the size of the experimental design increases. The sparse PCE models constructed by OMP lead to poor prediction performances with $\bar{Q}_{\text{test}}^2<0.16$ while LARS is a better choice to built the sparse PCE model in this case study. While the inner CV always assesses the PCE models in an overoptimistic way, the assessment by outer CV is close to what is obtained by independent testing. Moreover, the outer CV based on LOOCV slightly underestimates the determination coefficient for most cases meaning that it tends to give a conservative error estimate. Then, the whole set of data is put in the training (including outer CV) process and the obtained $Q^2_{\text{OCV}}=0.6812$ with LARS, which is smaller than $Q_{\text{ICV}}^2$ in Fig.~\ref{Prediction_LARS_fullData}. { One can also observe that all values of $\bar{Q}_{\text{ICV}}$ are insensitive to the size of training dataset with OMP and close to $1$, which is much larger than the reference value. These significant biases may be explained by observing from Fig. \ref{Coeff_FullSet} that OMP tends to yield a very flexible model, i.e., with a large value of $p$ and $j$ in the training process (see Algorithm \ref{alg:OMP}). Since a large number of candidate PCE models are searched for in the model selection, the inner CV has a high possibility to yield an over-optimistic model assessment. As a result, the average value of the determination coefficient is close to $1$.}  

\subsection{Data preprocessing and refined surrogate model based on sensitivity analysis}
\begin{figure}[!h]
	\centering
	\includegraphics[width=0.3\linewidth]{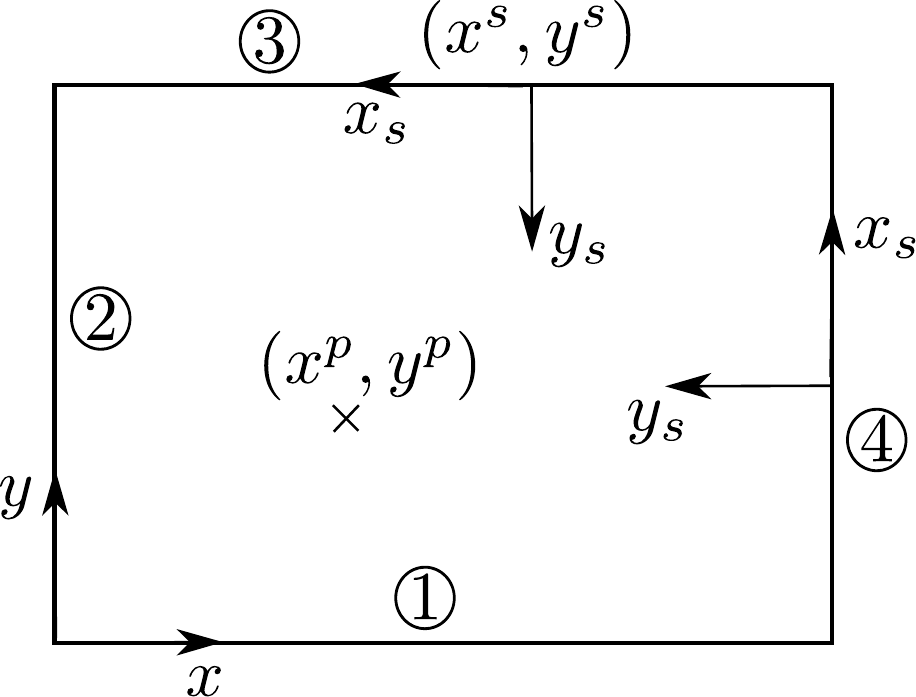}
	\caption{Specific absorption rate - formulation of input uncertainties in the local polar coordinate system of the WiFi box.}
	\label{fig:dataReformulation}
\end{figure}
The reflections by walls, ground and ceiling are ignored for the computation of SAR. Therefore, the SAR distribution is determined by the {\it relative position and orientation} of the human model vs. the WiFi box. Then, the considered input uncertainties can be represented in the local coordinate system of the WiFi box, as sketched in Fig.~\ref{fig:dataReformulation}, where the $y$ axis of the local coordinate system, denoted by $y_s$, is always orthogonal to the box surface and directed towards the inner space. As a result, the positive $y_s$ axis is opposite to the $y$ axis ($x$ axis) when the box is placed on the $3$rd wall ($4$th wall). Denote $(x^s,y^s)$, $(x^p,y^p)$ as Cartesian coordinates of the box and human model in the $x,y$ plane of the global coordinate system. Keeping $z^s$ to represent the height of the box, the other five uncertain variables can be reduced to three through the following formulas:
\begin{equation}
\begin{aligned}
&x_s^p = x^p - x^s, y_s^p = y^p - y^s, \theta_s^p = \theta^p, \, \text{if the box}\, \text{on the}\, 1\text{st wall,}\\
&x_s^p = y^s - y^p, y_s^p = x^p - x^s, \theta_s^p = (\theta^p+90^{\circ}) \, \text{mod.} \, (360^\circ), \, \text{if the box}\, \text{on the}\, 2\text{nd wall,}\\
&x_s^p = x^s - x^p, y_s^p = y^s - y^p, \theta_s^p = (\theta^p+180^{\circ}) \, \text{mod.} \, (360^\circ), \, \text{if the box}\, \text{on the}\, 3\text{rd wall,}\\
&x_s^p = y^p - y^s, y_s^p = x^s - x^p, \theta_s^p = (\theta^p+270^{\circ})\, \text{mod.} \, (360^\circ), \, \text{if the box}\, \text{on the}\, 4\text{th wall,}
\end{aligned} 
\end{equation} 
where ``mod." denotes the modulo operator. $(r_s^p,\psi_s^p)$, as the polar coordinates of $(x_s^p,y_s^p)$, denote the relative distance and angle, respectively. Thus, four variables $r_s^p$, $\psi_s^p$, $\theta_s^p$, $z^s$ can fully represent the uncertainty of inputs.

\begin{figure}[!ht]
	\centering
	\subfigure[LARS, $R^2=0.9659$]{\includegraphics[width = 0.33\linewidth]{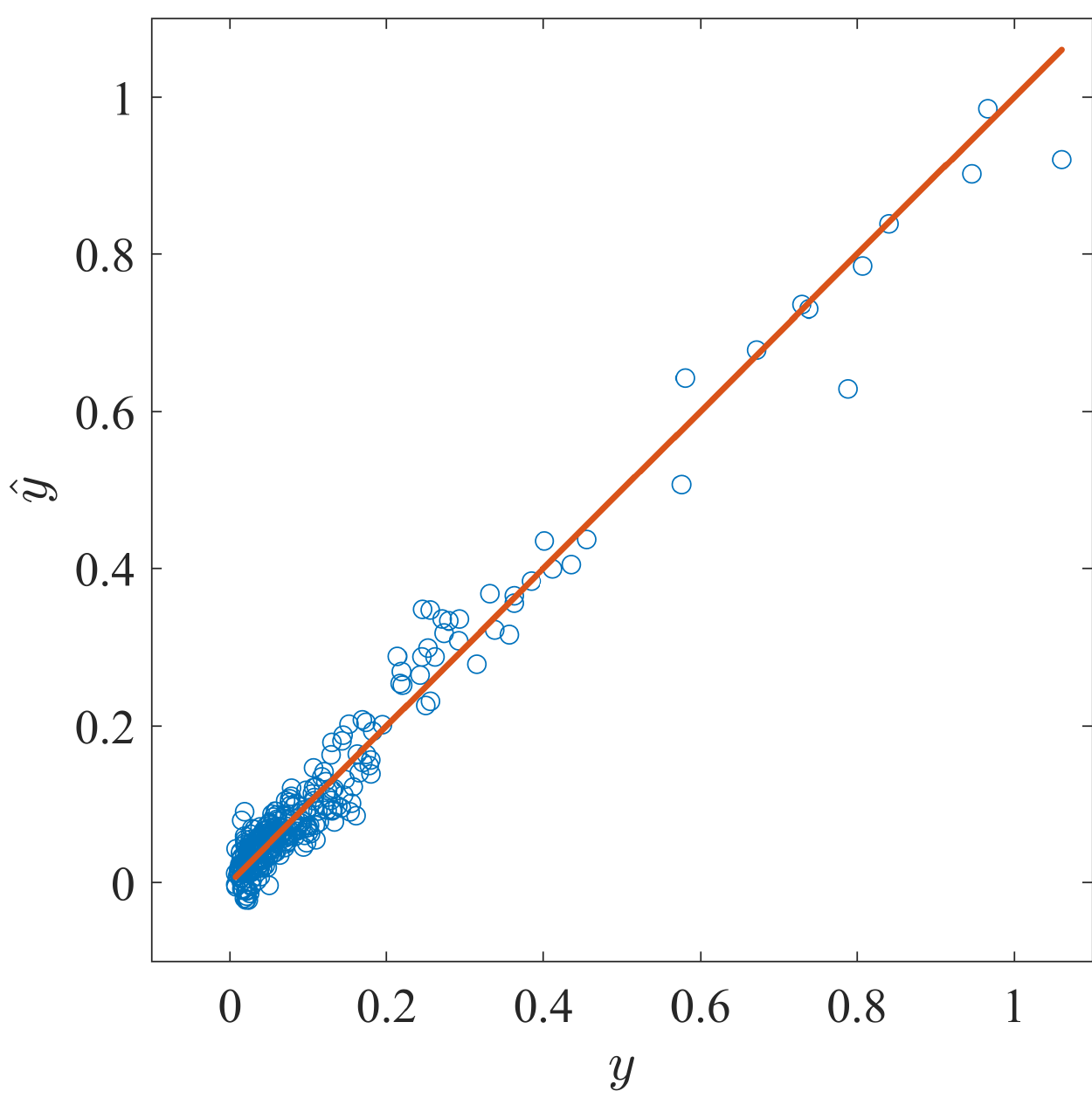}\label{Prediction_LARS_fullData_FourInput}}~
	\subfigure[OMP, $R^2=0.9915$]{\includegraphics[width = 0.33\linewidth]{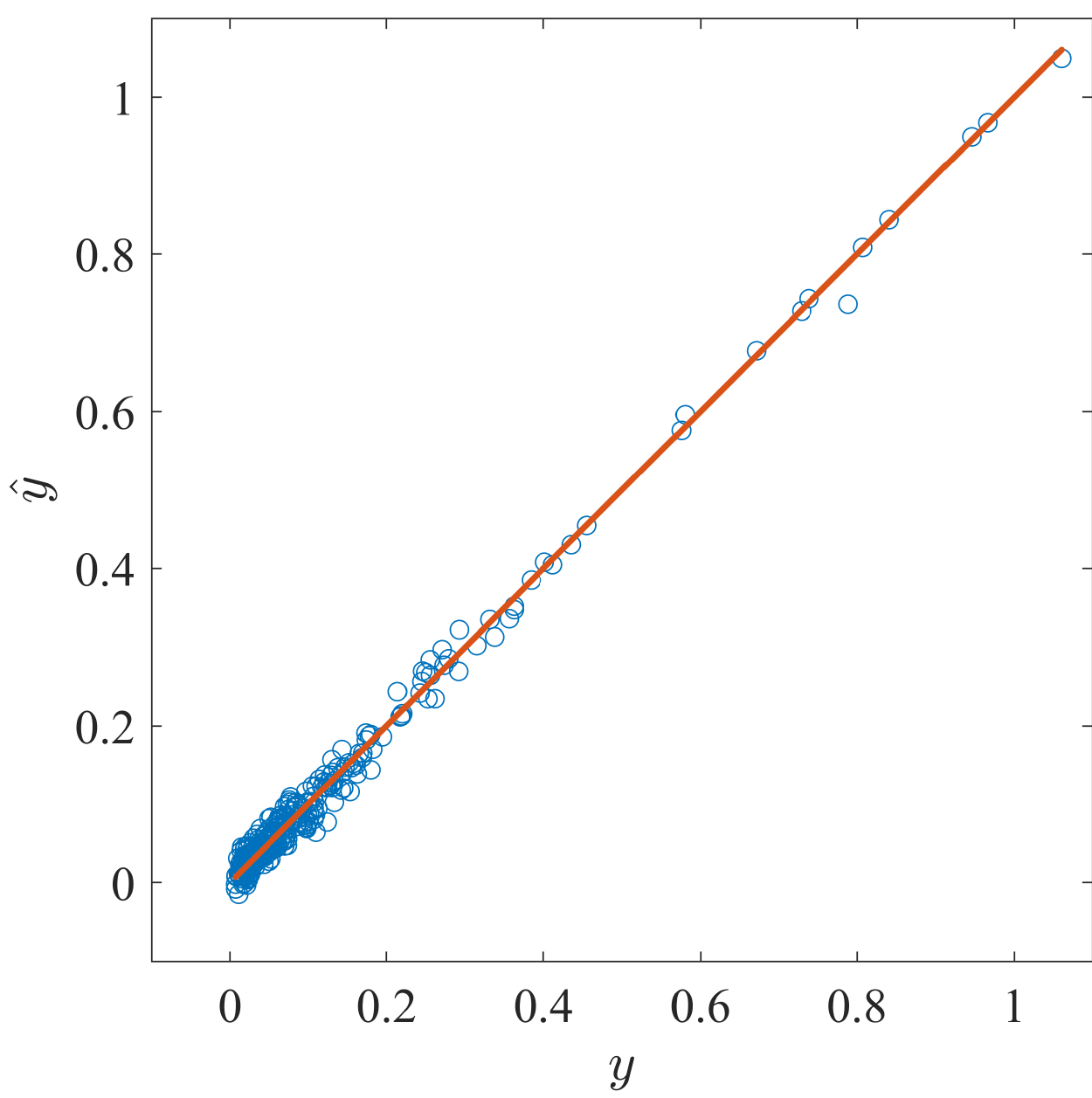}\label{Prediction_OMP_fullData_FourInput}}~
	\subfigure[Full model, $R^2=0.8811$]{\includegraphics[width = 0.33\linewidth]{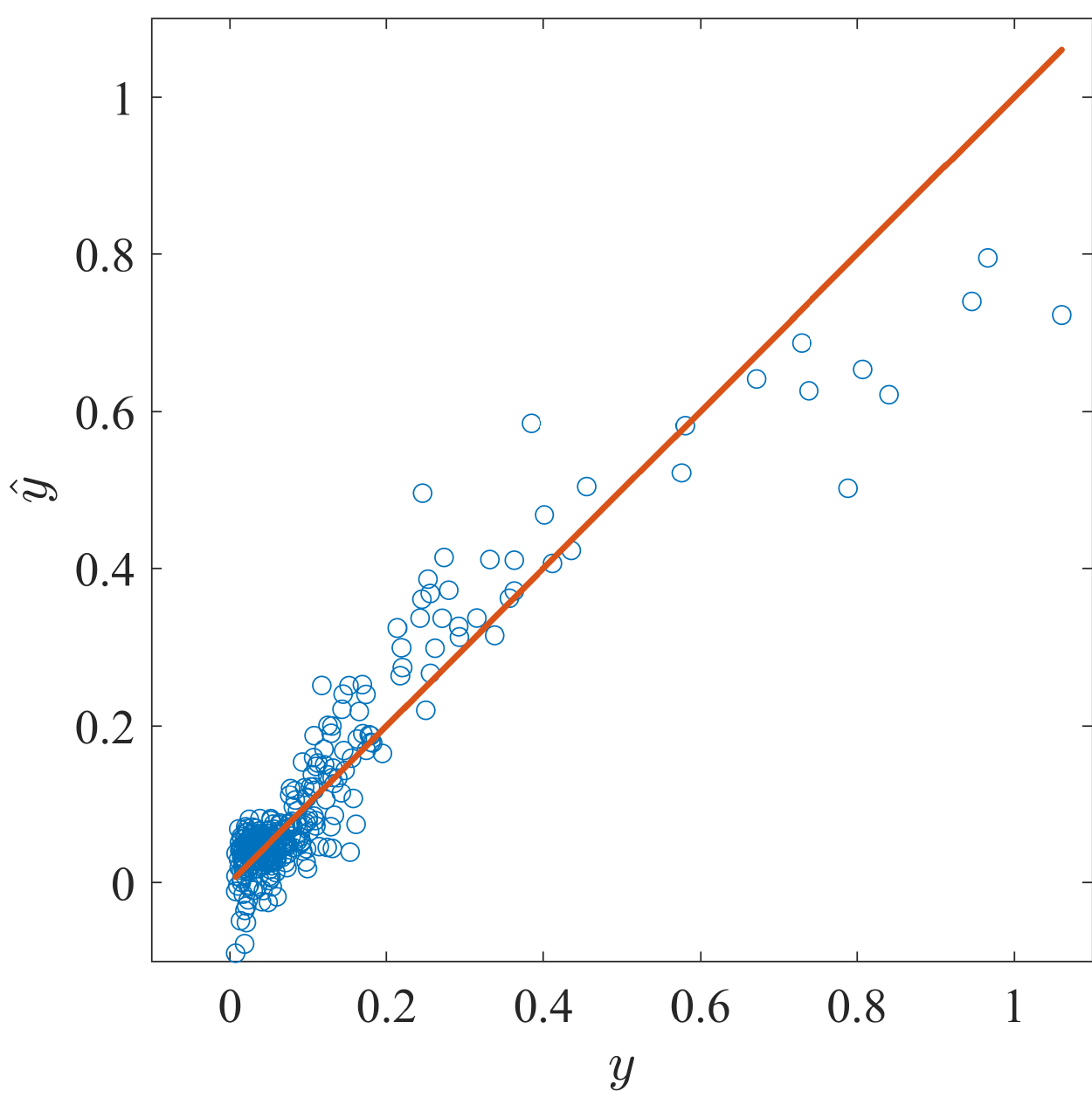}\label{Prediction_OLS_fullData_FourInput}}
	\caption{Specific absorption rate - prediction of training data with the constructed PCE model based on the reformulated data.}
	\label{Prediction_fullData_FourInput}
\end{figure}
\begin{figure}[!ht]
	\centering
	\subfigure[LARS, $30$ nonzero coefficients ]{\includegraphics[width = 0.38\linewidth]{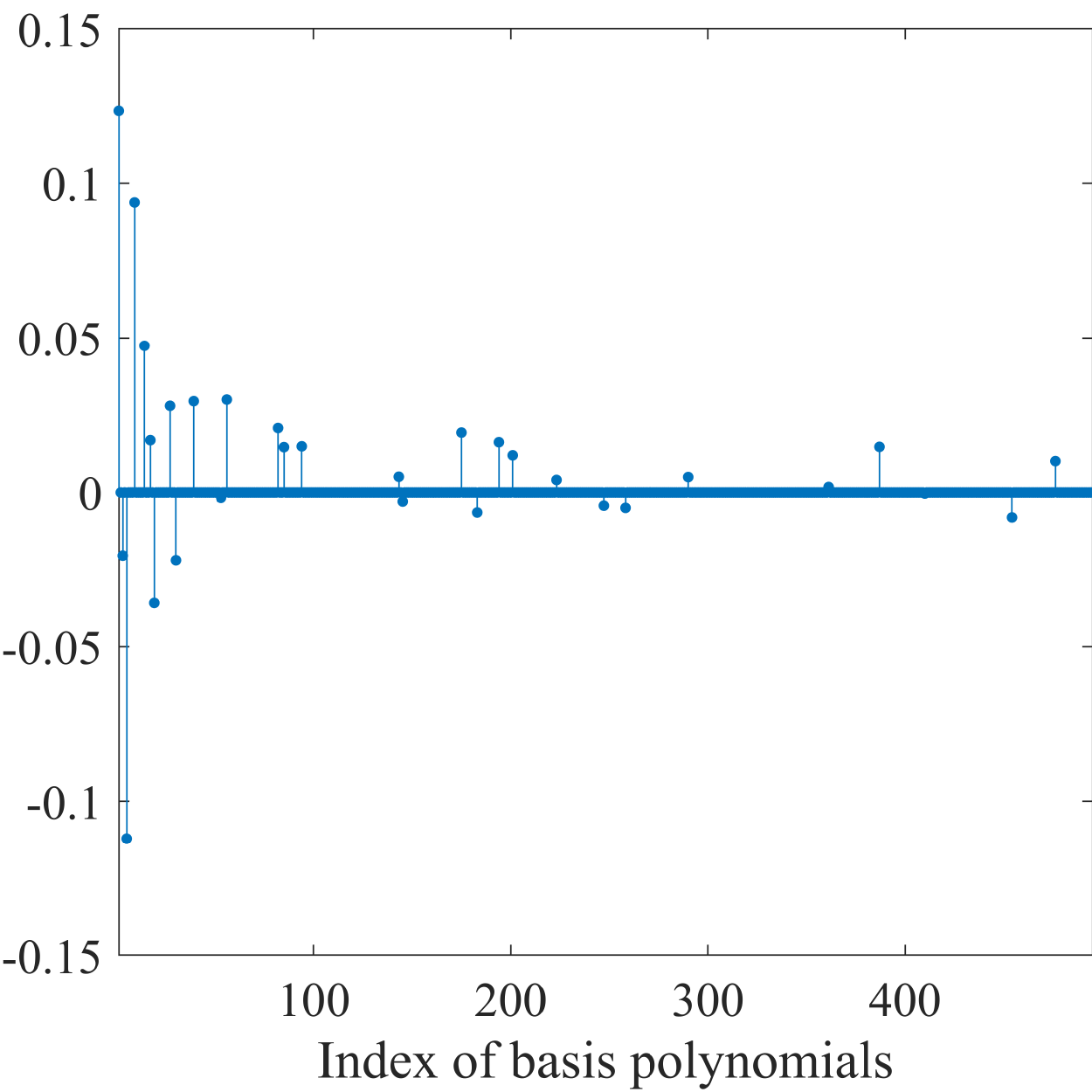}\label{Coeff_LARS_fullData_FourInput}}~
	\subfigure[OMP, $71$ nonzero coefficients]{\includegraphics[width = 0.38\linewidth]{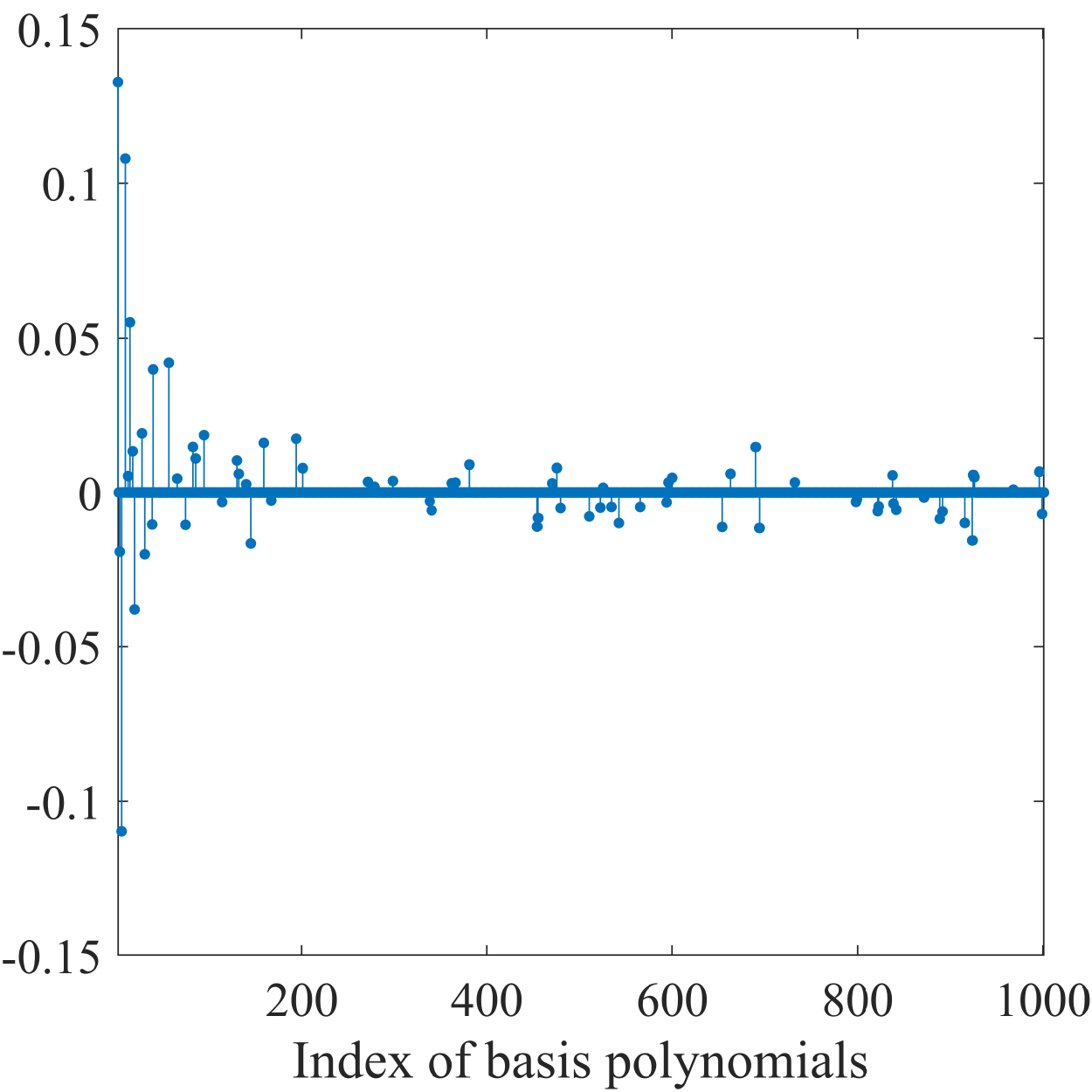}\label{Coeff_OMP_fullData_FourInput}}
	\caption{Specific absorption rate - expansion coefficients of the constructed sparse PCE models with the reformulated data.}
	\label{Coeff_FullSet_FourInput}
\end{figure}

{With the whole set of the reformulated data, the prediction of the training dataset is shown in Fig.~\ref{Prediction_fullData_FourInput}. While the well approximated prediction is obtained with LARS and OMP, the performance of the full model significantly improves (relative to the performance in Fig.~\ref{Prediction_fullData}\subref{Prediction_OLS_fullData}). The associated $R^2$ is increased from $0.4718$ to $0.8811$. The sparsity of the expansion coefficients of the sparse PCE models is illustrated in Fig.~\ref{Coeff_FullSet_FourInput}. $30$ out of $495$ polynomials are selected with LARS and $71$ out of $1001$ with OMP. Remark that large indices indicate high-degree polynomials and the amplitude of the coefficient is proportional to the influence of the corresponding polynomial. As seen, most of influential polynomials are with a small degree.}
\begin{figure}[!ht]
	\centering
	\subfigure[LARS]{\includegraphics[width = 0.5\linewidth]{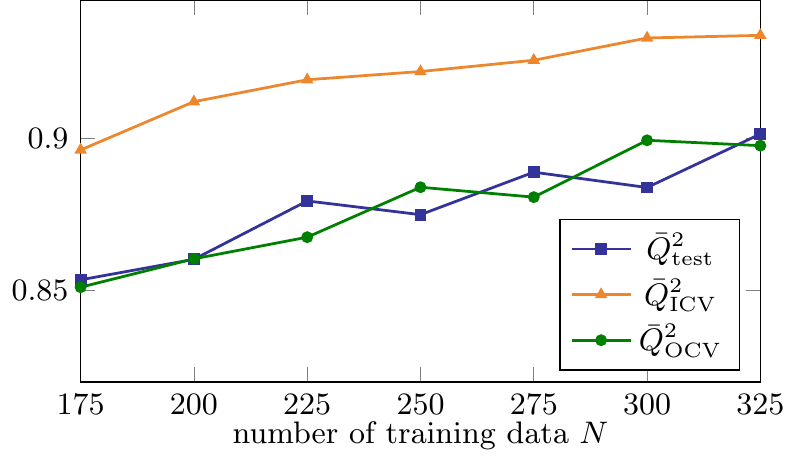}\label{CMV_LOO_LARS_maxOrder10_4Input}}~
	\subfigure[OMP]{\includegraphics[width = 0.5\linewidth]{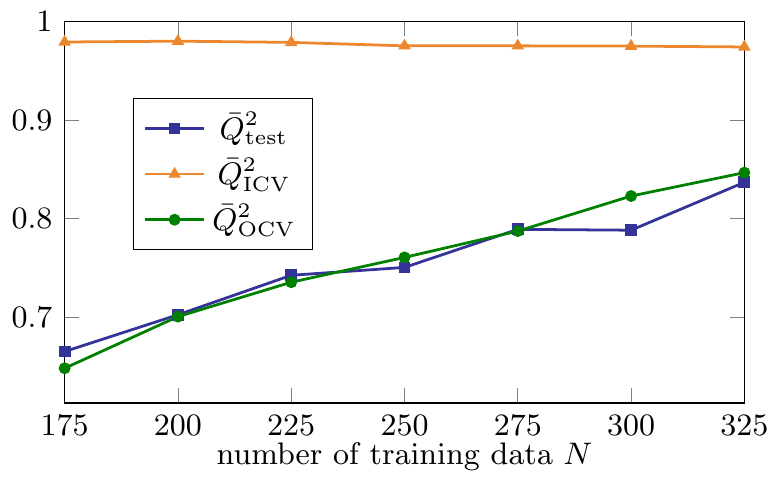}\label{CMV_LOO_OMP_maxOrder10_4Input}}
	\caption{Specific absorption rate - mean of $Q^2$ over $100$ replications versus $N$ with the reformulated data.}
	\label{Prediction_CMV_LOO_4Input}
\end{figure}

Performing outer CV by LOOCV, the accuracy of model assessment is checked by independent testings. Varying the size of training data (the size of the testing data as well), the averaged determination coefficients over $100$ replications are shown in Fig.~\ref{Prediction_CMV_LOO_4Input}. Model assessment by inner CV reveals optimistic (especially with OMP) whereas the outer CV shows unbiasedness. Based on the whole set of data, the outer CV yields $Q^2_{\text{OCV}}=0.9024$ with LARS and $0.8337$ with OMP, which are much larger than the values obtained with six input variables. { This can be explained by the observation that the number of nonzero expansion coefficients, i.e., the number of selected basis polynomials, is reduced. As the required number of samples is proportional to the size of basis, the modeling performance is improved with the same experimental design.}

\begin{table}[!h]
	\centering
	\begin{tabular}{l|*{4}{c}|c}
		\toprule
		& $r_s^p$ & $\psi_s^p$ & $z^s$ & $\theta_s^p$ & {Sum}\\
		\midrule
		LARS & 0.9772 & 0.0405 & 0.2039 & 0.0404 & {1.2620}\\
		OMP & 0.9773 & 0.0677 & 0.2324 & 0.0555 & {1.3329}\\
		\bottomrule
	\end{tabular}
	\caption{Specific absorption rate - total Sobol' indices computed with the PCE model based on the whole set of reformulated data.}
	\label{tab:totalSobol}
\end{table}

Global sensitivity analysis is performed through the computation of total Sobol' indices, which measure the total contribution (summation of individual and interacted ones) of an input variable to the response uncertainty. Based on the PCE model constructed with the whole set of data, the total Sobol' indices are computed from PCE coefficients \cite{sudret2008global} and provided in Table~\ref{tab:totalSobol}. The PCE models trained with LARS and OMP support the same conclusion that the uncertainty of response is mainly contributed to by $r_s^p$ and $z^s$ and the effects of $\psi_s^p$ and $\theta_s^p$ can be neglected. {The sum of total Sobol' indices is larger than $1$ for both models indicating that the interaction between input variables matters. Furthermore, the average value of the second-order Sobol' indices related with $r_s^p$ and $z^s$ is $0.1584$ with LARS and $0.2077$ with OMP, which indicate that $z^s$ impacts on the output mainly via its interaction with $r_s^p$.} 

\begin{figure}[!ht]
	\centering
	\subfigure[LARS, $R^2=0.9601$]{\includegraphics[width = 0.33\linewidth]{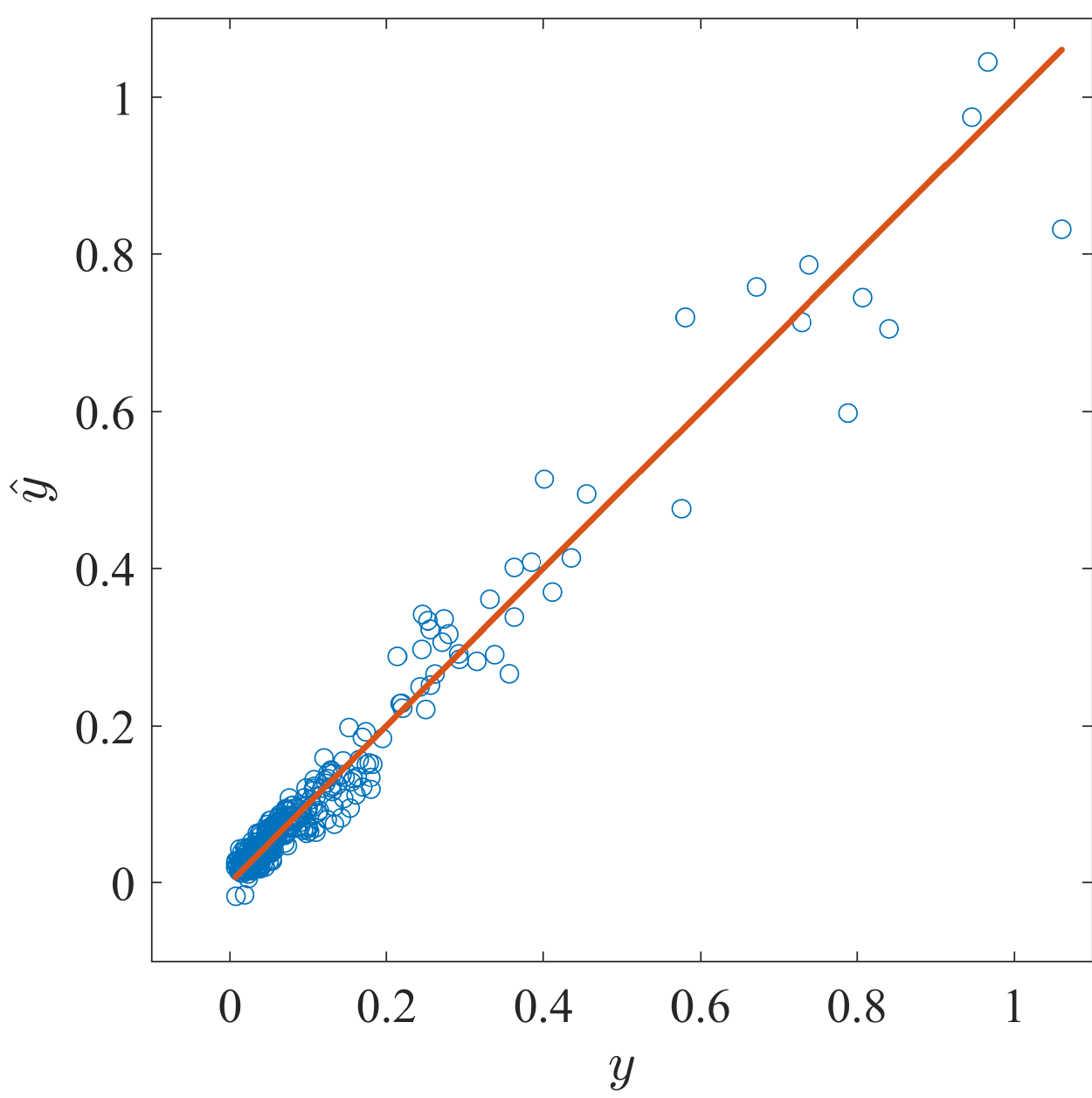}\label{Prediction_LARS_fullData_TwoInput}}~
	\subfigure[OMP, $R^2=0.9629$]{\includegraphics[width = 0.33\linewidth]{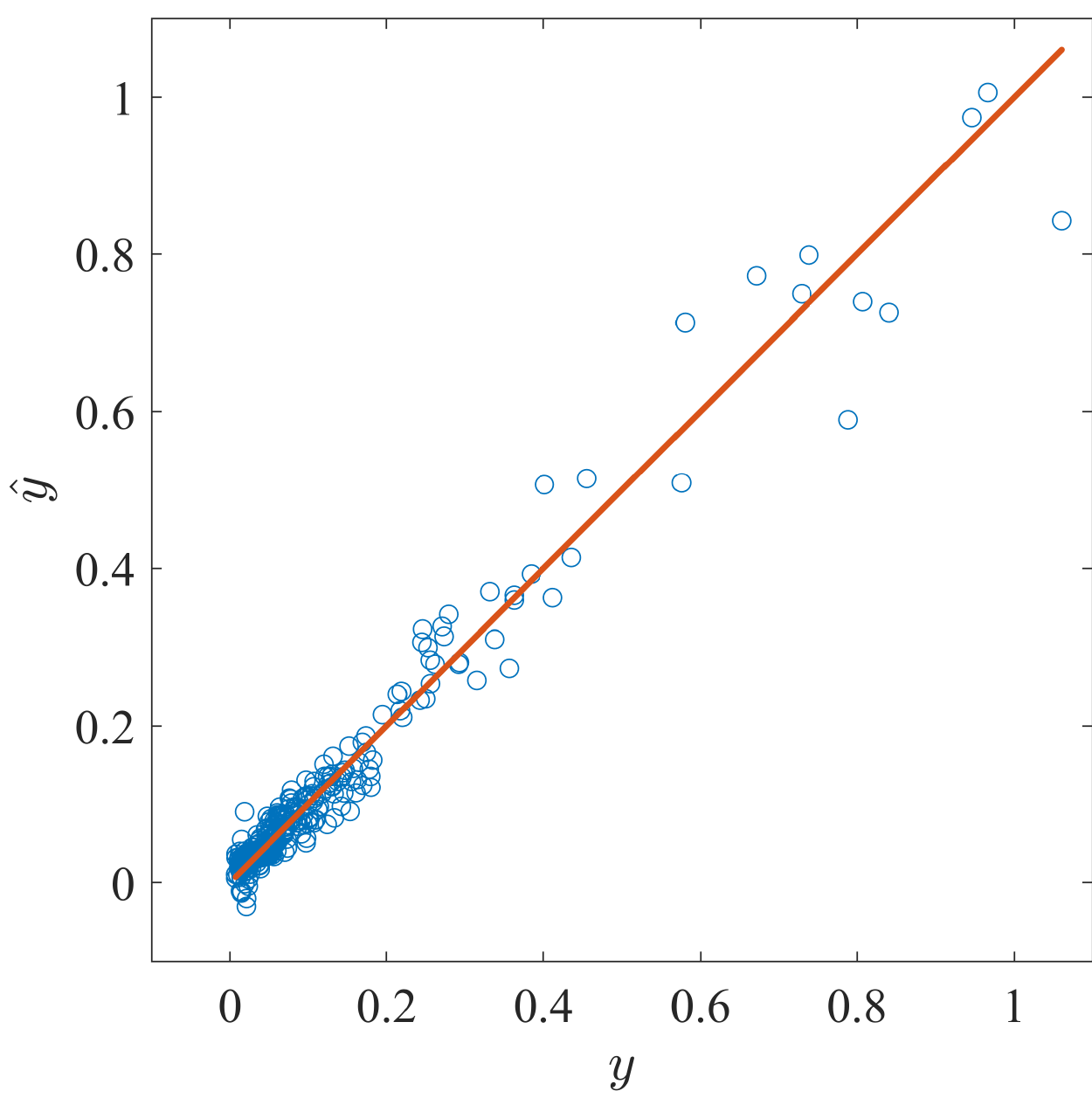}\label{Prediction_OMP_fullData_TwoInput}}~
	\subfigure[Full model, $R^2=0.9549$]{\includegraphics[width = 0.33\linewidth]{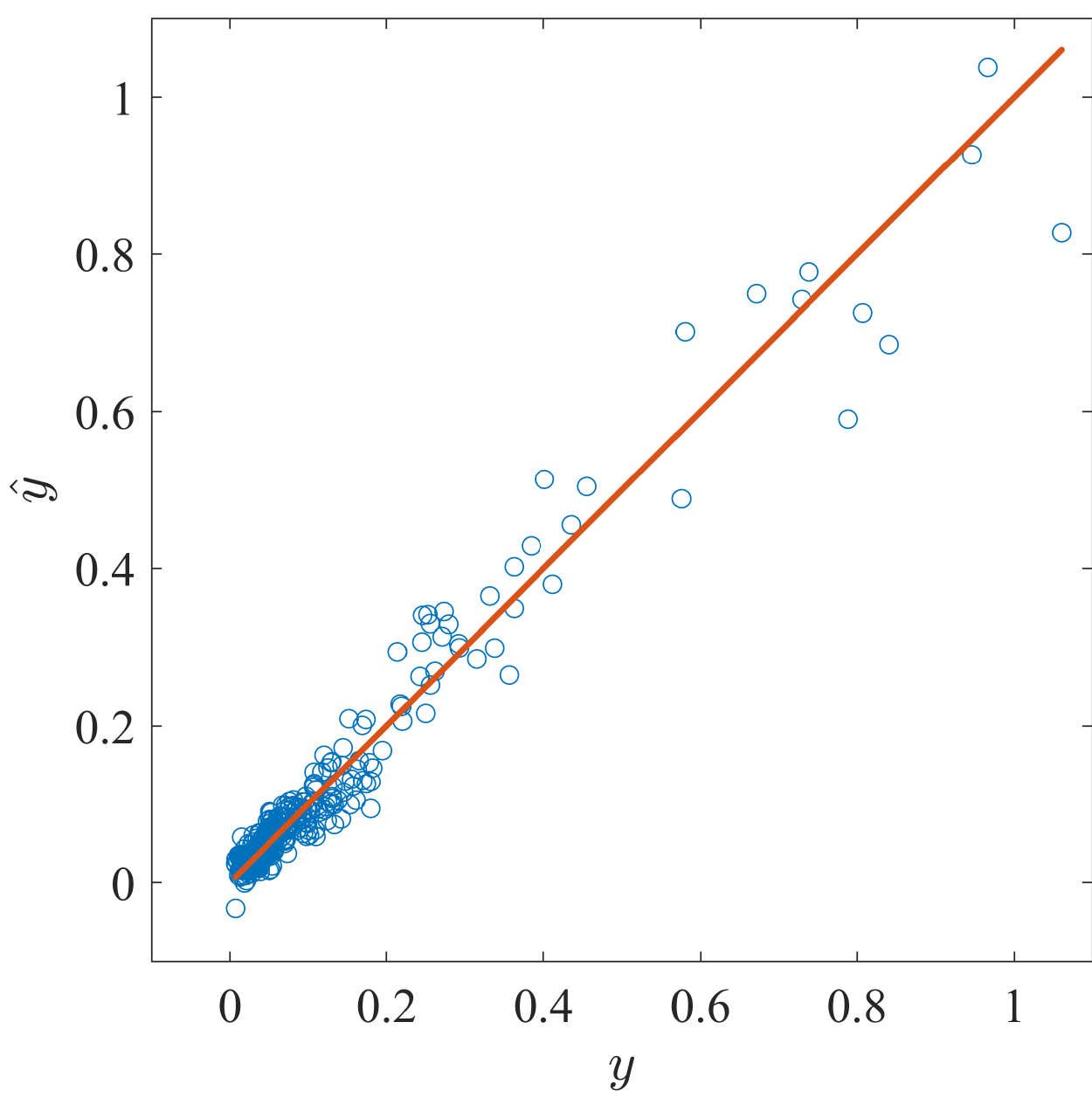}\label{Prediction_OLS_fullData_TwoInput}}
	\caption{Specific absorption rate - prediction of training data with the PCE model when only variables $r_s^p$ and $z^s$ are considered.}
	\label{Prediction_fullData_TwoInput}
\end{figure}
\begin{figure}[!ht]
	\centering
	\subfigure[LARS, $23$ nonzero coefficients ]{\includegraphics[width = 0.4\linewidth]{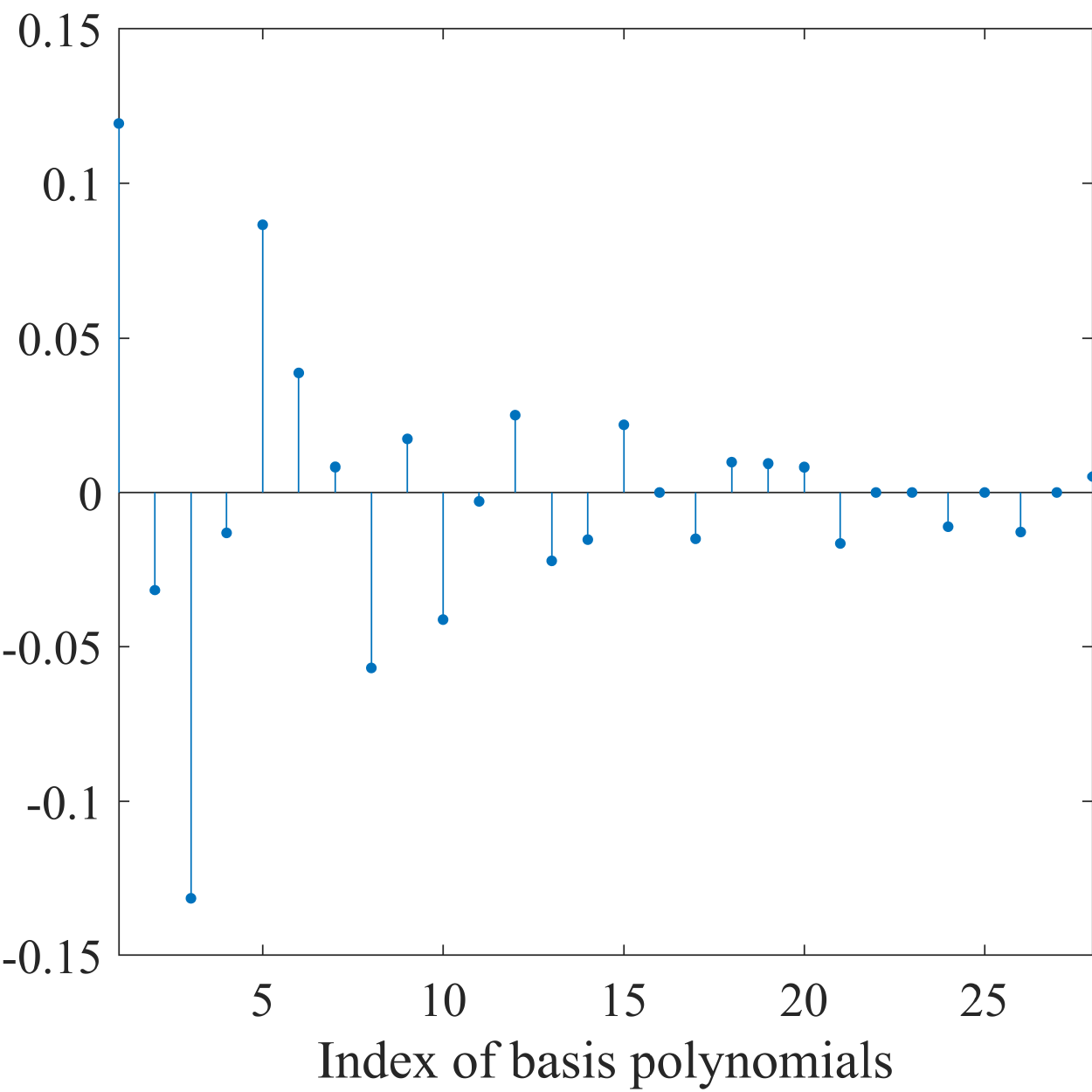}\label{Coeff_LARS_fullData_TwoInput}}~
	\subfigure[OMP, $24$ nonzero coefficients]{\includegraphics[width = 0.4\linewidth]{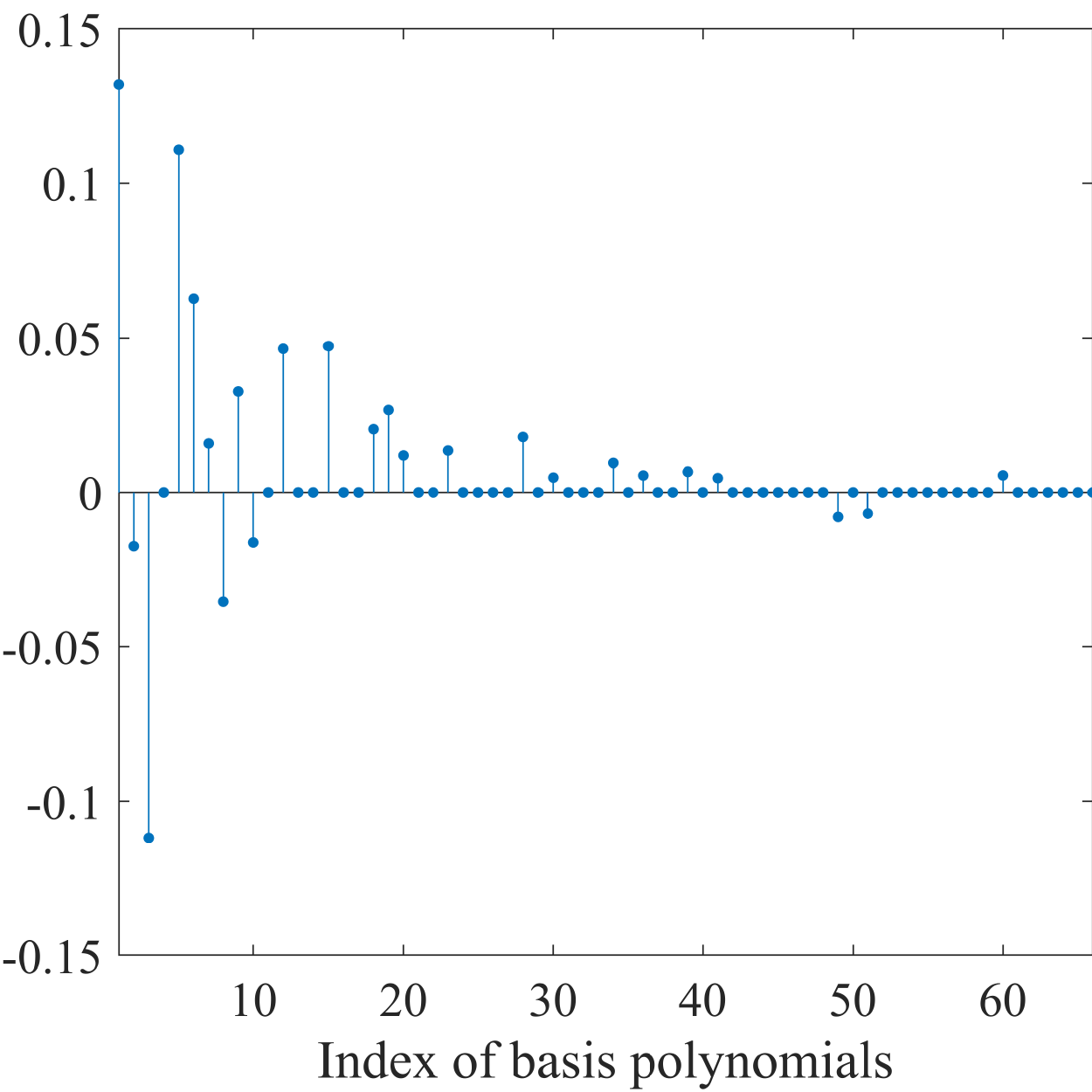}\label{Coeff_OMP_fullData_TwoInput}}
	\caption{Specific absorption rate - expansion coefficients of the constructed sparse PCE models when only variables $r_s^p$ and $z^s$ are considered.}
	\label{Coeff_FullSet_TwoInput}
\end{figure}
Following the result of the sensitivity analysis, surrogate models are constructed only considering two uncertain variables, $r_s^p$ and $z^s$. {The training data are well predicted by the constructed PCE model with the three approaches. Since the number of input variables is small, the curse-of-dimensionality is avoided and the performance of the full model gets improved. As shown by Fig.~\ref{Coeff_FullSet_TwoInput}, the involved coefficients of the sparse PCE models are not sparse considering that 23 out of 28 coefficients are nonzero with LARS and 24 out of 66 with OMP.}

\begin{figure}[!ht]
	\centering
	\subfigure[LARS]{\includegraphics[width = 0.5\linewidth]{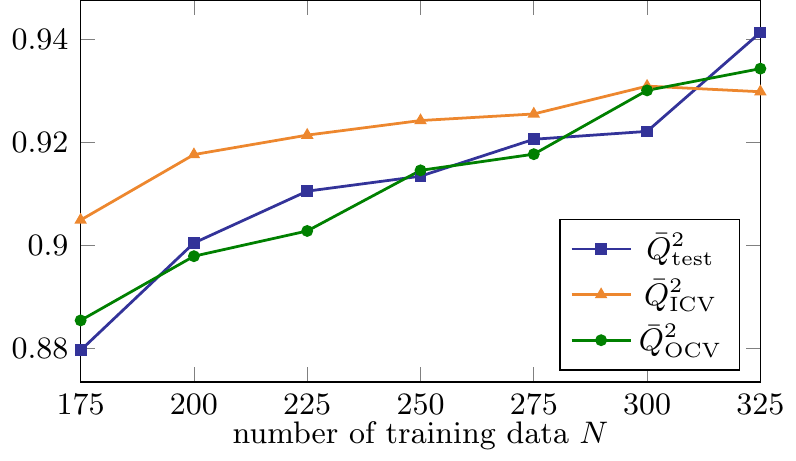}\label{CMV_LOO_LARS_maxOrder10_2Input}}~
	\subfigure[OMP]{\includegraphics[width = 0.5\linewidth]{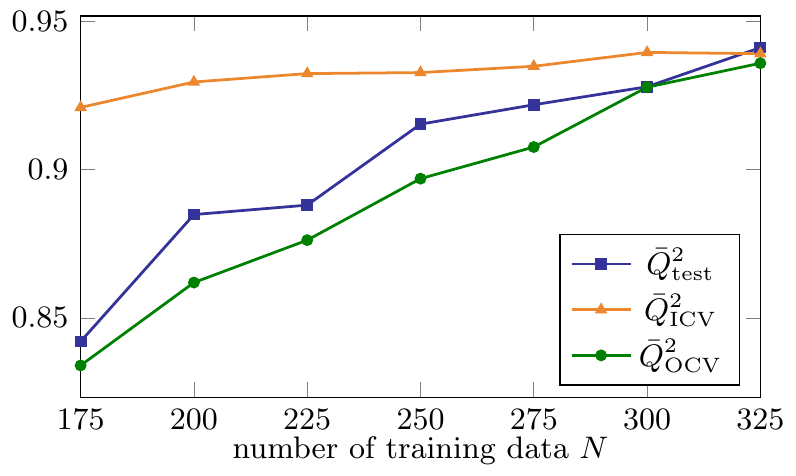}\label{CMV_LOO_OMP_maxOrder10_2Input}}
	\caption{Specific absorption rate - mean of $Q^2$ over $100$ replications versus $N$ with surrogate models only considering variables $r_s^p$ and $z^s$.}
	\label{Prediction_CMV_LOO_2Input}
\end{figure}
\begin{figure}[!ht]
	\centering
	\subfigure[LARS, $\bar{Q}^2_{\text{OCV}}=0.9461$]{\includegraphics[width = 0.32\linewidth]{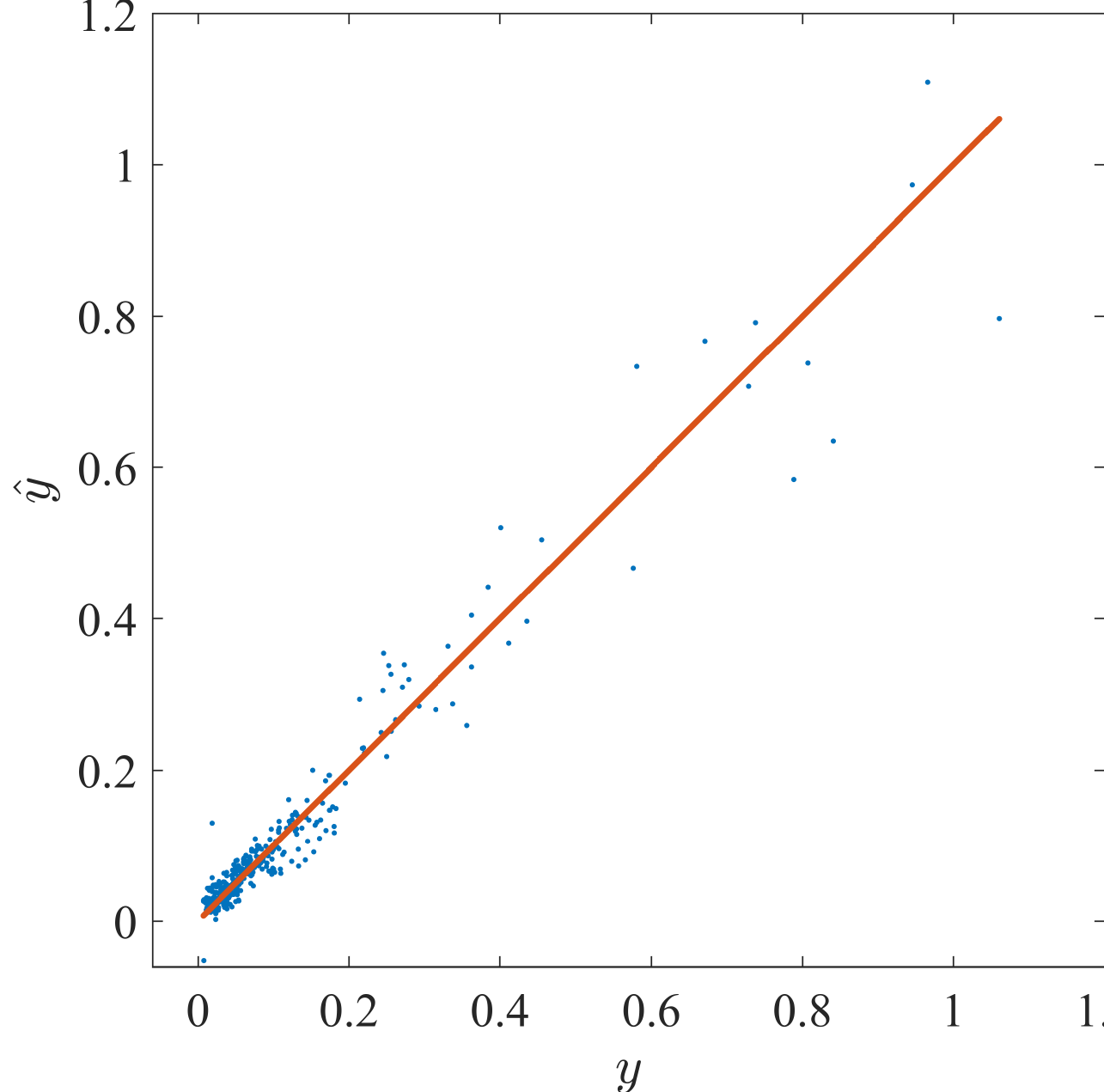}\label{Prediction_CMV_LARS_2Input}}
	\subfigure[OMP, $\bar{Q}^2_{\text{OCV}}=0.9448$]{\includegraphics[width = 0.32\linewidth]{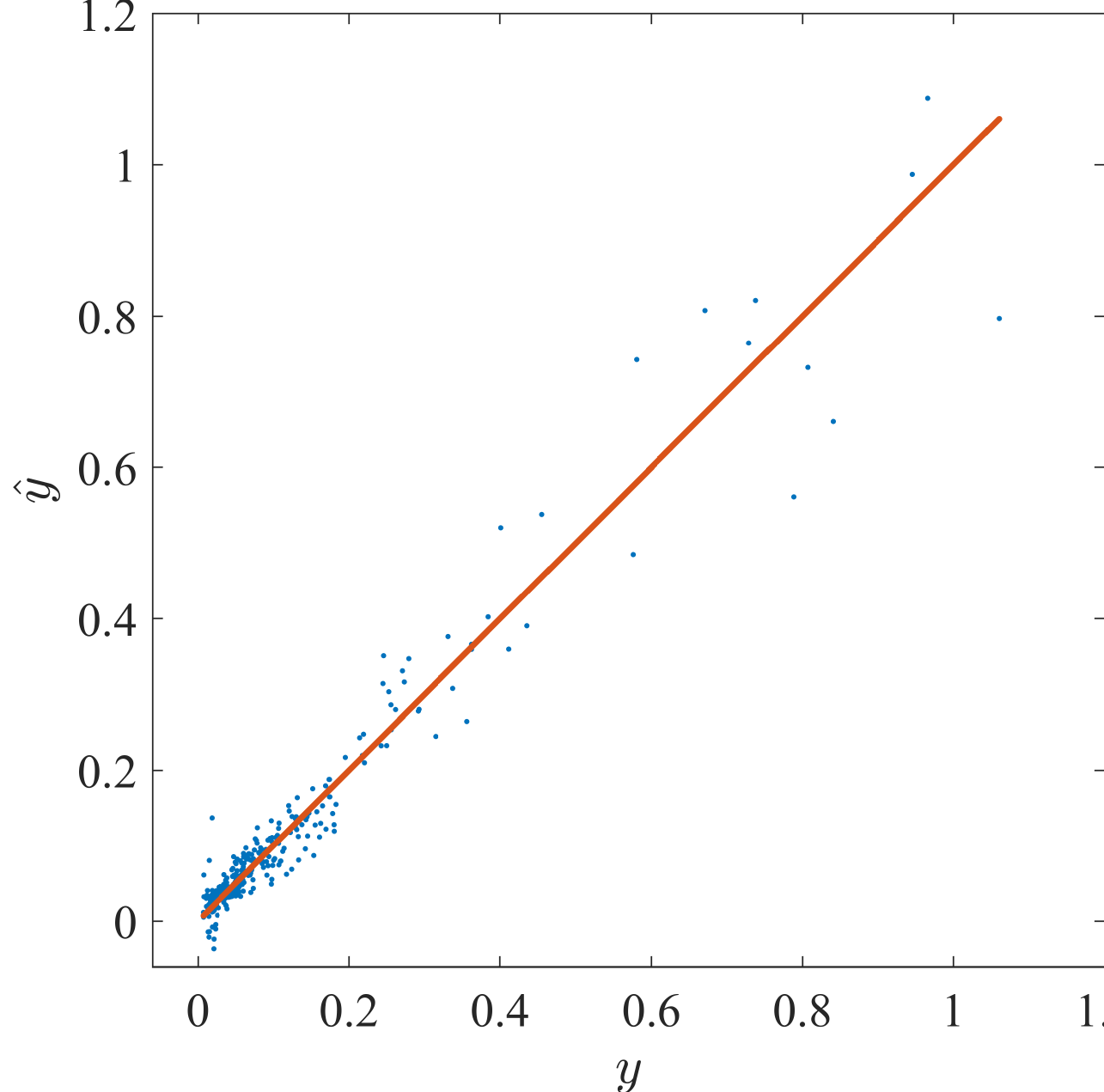}\label{Prediction_CMV_OMP_2Input}}
	\subfigure[Full model, $\bar{Q}^2_{\text{OCV}}=0.9405$]{\includegraphics[width = 0.32\linewidth]{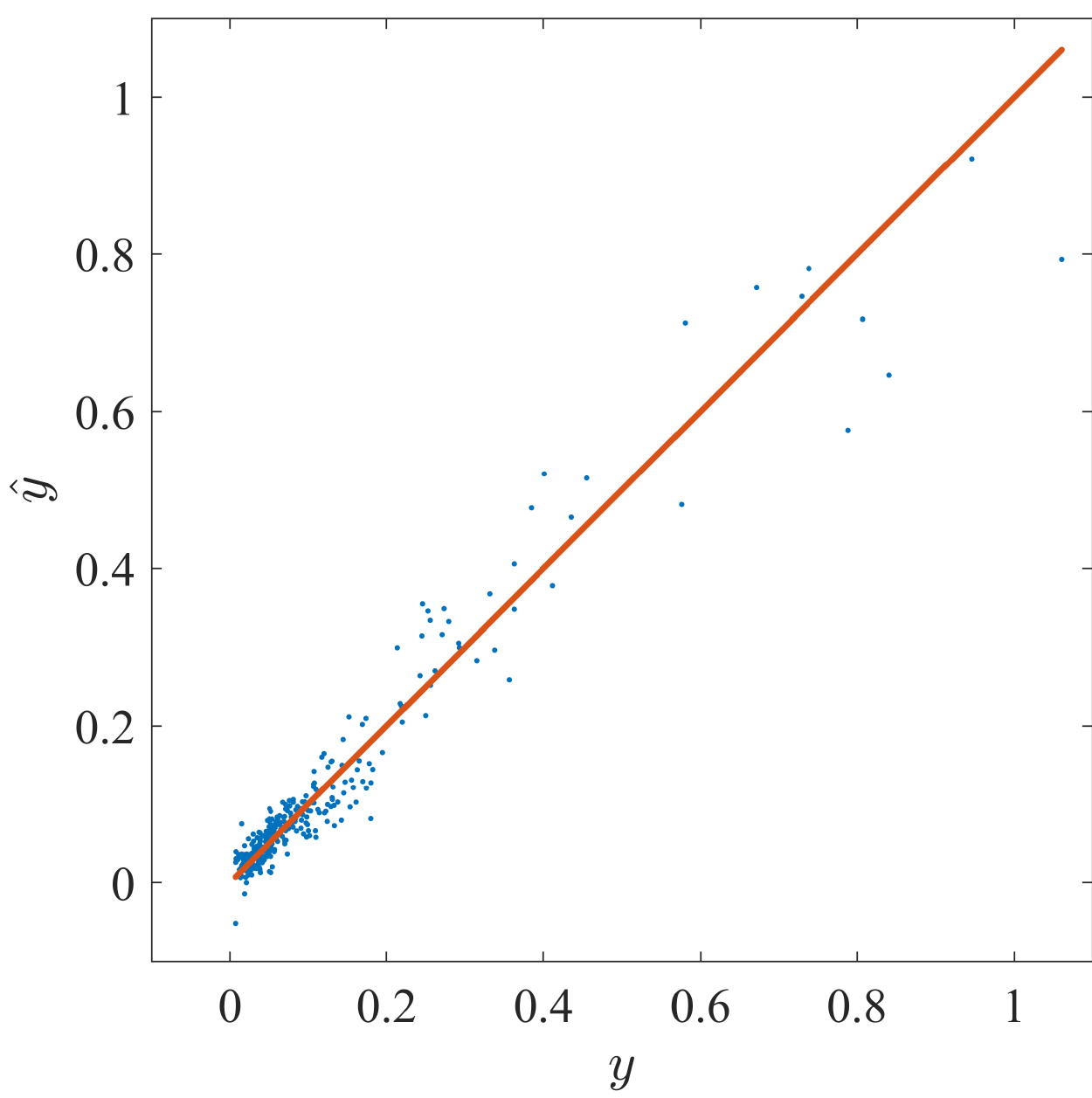}\label{diagSARLOO_OLS_2Input}}
	\caption{Specific absorption rate - outer cross validation by leaving one sample out when only $r_s^p$ and $z^s$ are considered as uncertain variables.}
	\label{Prediction_CMV_2Input}
\end{figure}

The accuracy of model assessment by inner and outer CV is checked by independent testings, as shown in Fig.~\ref{Prediction_CMV_LOO_2Input}. With LARS, while inner CV underestimates the CV error for most cases, the deviation of assessments by inner CV is smaller than $0.03$. With OMP, the deviation for assessments by inner CV becomes larger (e.g. the deviation $\approx0.08$ when $N=175$) but outer CV still closely follows the reference value (provided by independent testings). Making use of the whole set of data to construct PCE models and running outer CV by leaving one sample out, the prediction of all samples left out is shown in Fig.~\ref{Prediction_CMV_2Input}, where unbiased estimations are observed with the three approaches. The associated determination coefficient equals $0.9461$ with LARS and $0.9448$ with OMP, which are larger than the values obtained with four input variables, as described before. {With only two input variables, the full model shows prediction performances similar to the ones obtained by sparse PCE.} 

\begin{figure}[!ht]
	\centering
	\begin{minipage}{.45\textwidth}
		\centering
		\includegraphics[width=.9\linewidth]{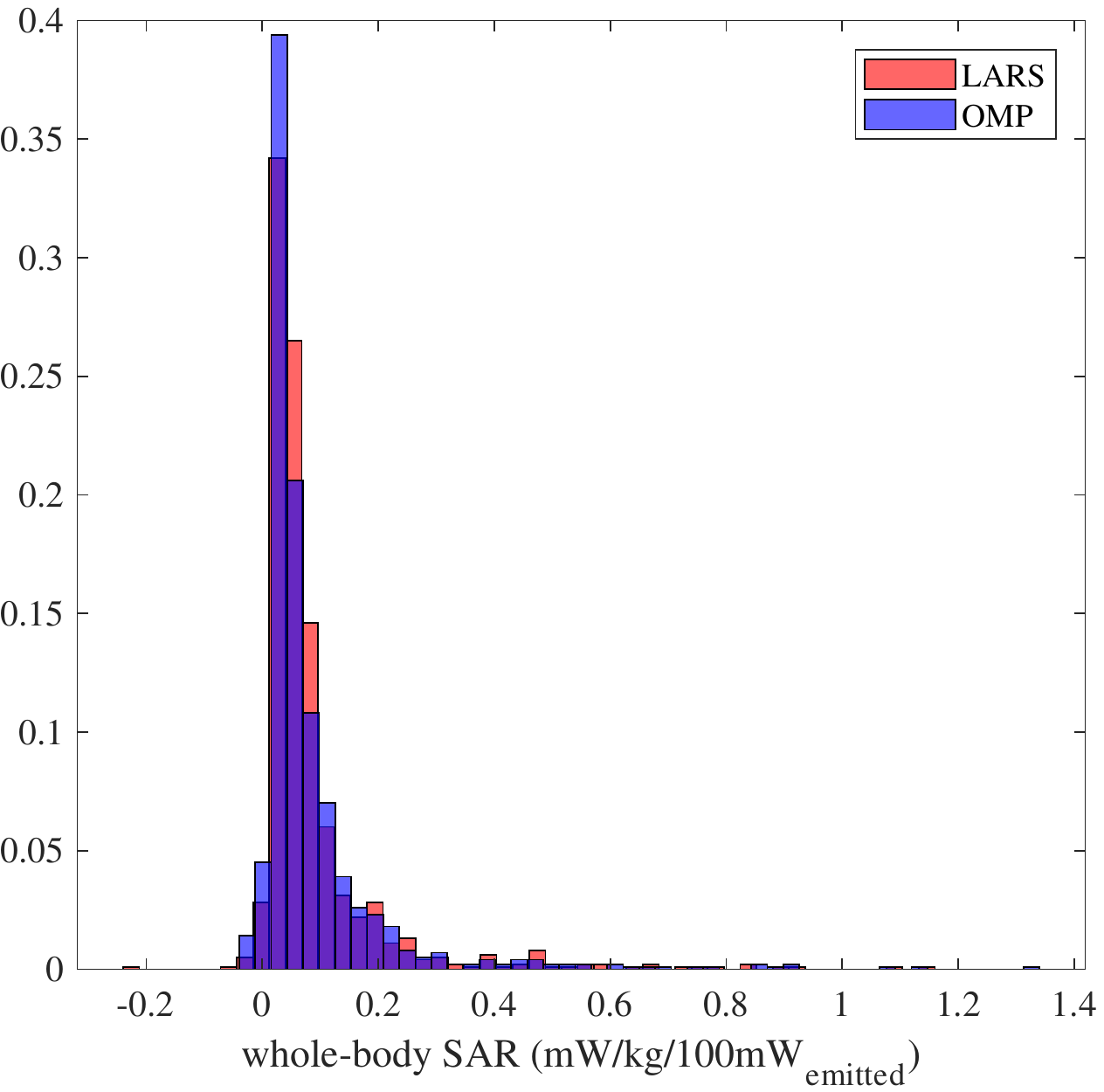}
		\caption{Normalized histogram of predicted SAR with the PCE model based on LARS and OMP, respectively.}
		\label{histogramSAR}
	\end{minipage}
	\begin{minipage}{.45\textwidth}
		\centering
		\includegraphics[width=.9\linewidth]{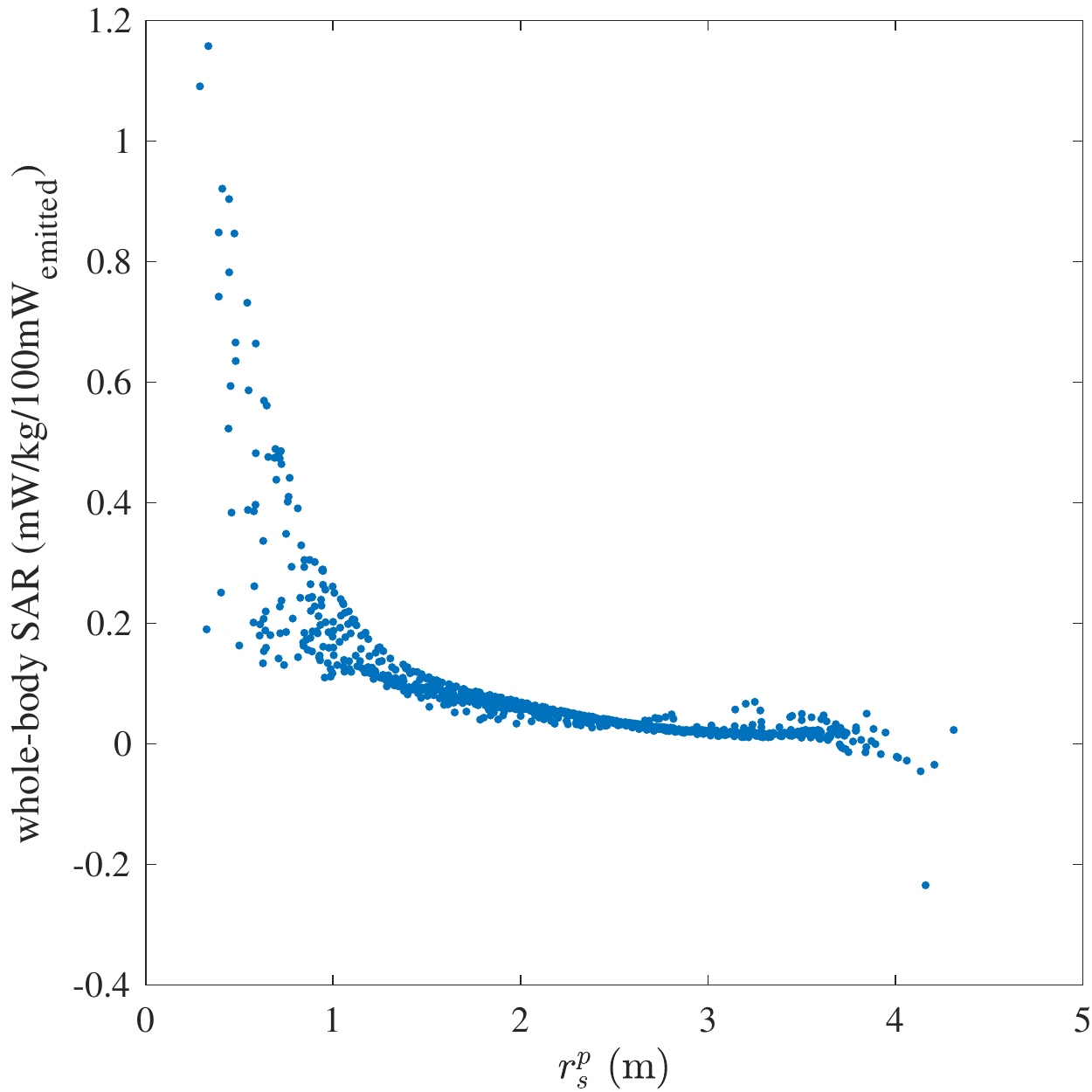}
		\caption{Predicted SAR versus the relative distance with the PCE model based on LARS.}
		\label{predSAR_LARS}
	\end{minipage}
\end{figure}
Constructing a PCE model with the whole set of data and considering only parameters $r_s^p$ and $z^s$, more insights may be achieved from the prediction of the $1,000$ unobserved inputs, which are obtained from the LHS method. To compare the prediction results by the PCE model with LARS and OMP, the normalized histograms (a.k.a. probability of occurrence) are given in Fig.~\ref{histogramSAR}. As observed, the SAR values predicted by the two models have similar distributions and most predictions are in the range of $[0,0.2]$. Remark that negative SAR values appear for a small portion of samples due to the approximation error of the surrogate models. This phenomenon usually happens when the true value is close to zero.   
	
The high correlation between the output and the distance $r_s^p$ is shown in Fig.~\ref{predSAR_LARS}. The whole-body SAR decreases rapidly with $r_s^p$ but the decreasing rate is slower with larger $r_s^p$. Moreover, it seems that the variance of the SAR distribution becomes smaller with larger $r_s^p$. That maybe explained by the fact that small $r_s^p$ are corresponding to very near regions, while the electromagnetic scattering between source and human model is so complex that other factors (e.g., source height $z^s$) need to be considered. As the distance $r_s^p$ is larger, $r_s^p$ will dominate the level of human exposure.

\begin{figure}[!ht]
	\centering
	\includegraphics[width=.45\linewidth]{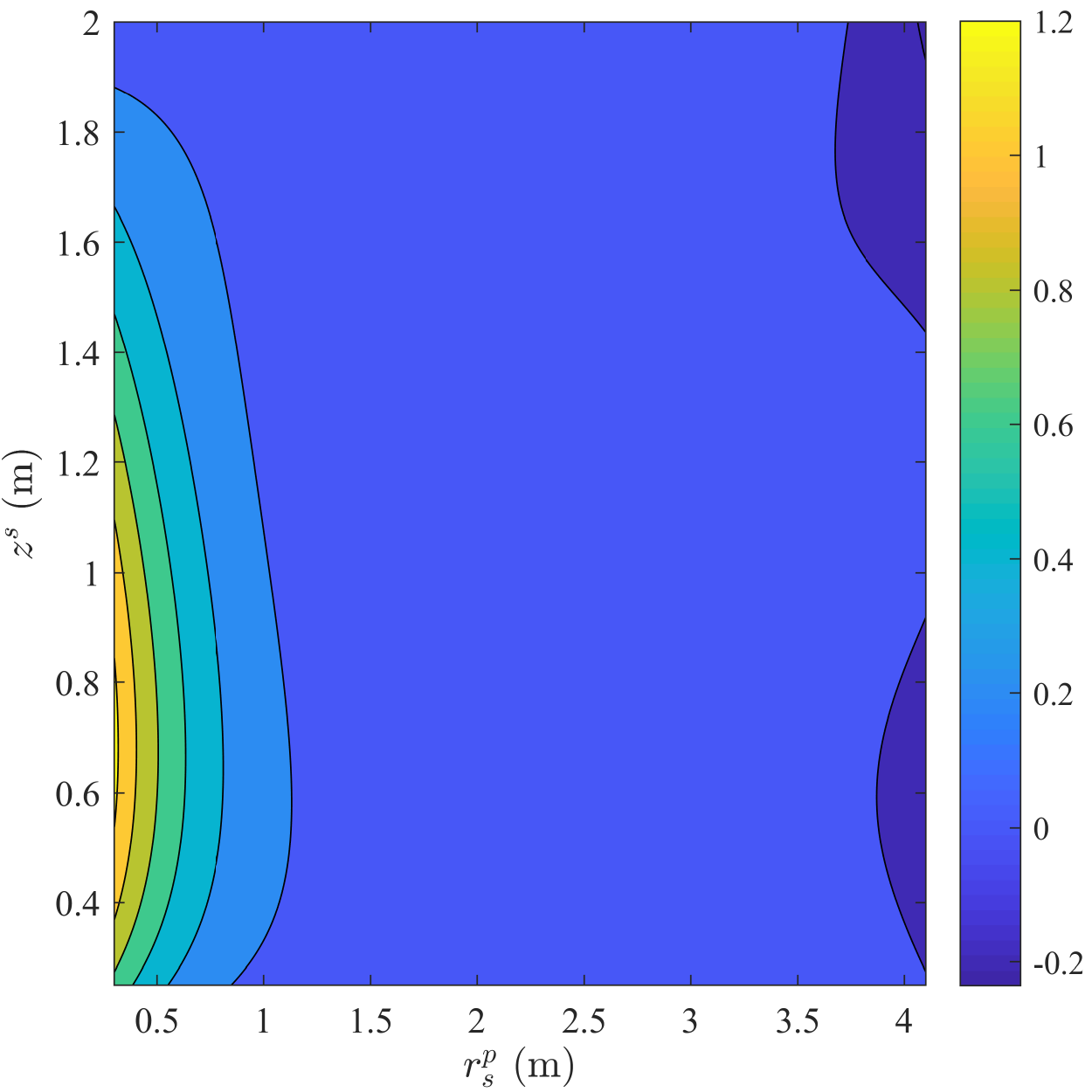}
	\caption{Contour of predicted SAR with varied $r_s^p$ and $z^s$.}
	\label{contourSAR}
\end{figure}
{The contour of the predicted SAR is given in Fig.~\ref{contourSAR} when $r_s^p$ and $z^s$ are varied. As observed, peak values of whole-body SAR happen when $r_s^p$ is small and $z^s$ is between $0.4$m and $1.1$m. Considering the height of human model equals $1.36$m, this influential region corresponds with the area where most of tissues locate.}

\section{Conclusions}
\label{sec:conclusions}

The human exposure by a WiFi box inside a room is statistically analyzed by constructing a surrogate model based on polynomial chaos expansion. While the samples of the input variables are determined by Latin hypercube sampling, the response is generated by combining measurements of scattered fields and an in-house FDTD code through the computation of a Huygens box. Based on a limited sample set, the sparse PCE model is constructed and assessed by cross validation (CV) (including inner CV and outer CV). After the pre-processing of the input variables and the sensitivity analysis, it is shown that the performance of the surrogate models can be improved. 

Inner CV is commonly used to perform model selection and assessment during the construction of sparse PCE models. However, since the CV error is optimized during the model selection, the model assessment by inner CV is found over-optimistic (especially when the surrogate model does not perform well due to a very small experimental design). The introduction of outer CV in the methodology of cross-model validation allows one to get more accurate estimation of the validation error.

Regarding the computational dosimetry example, the electromagnetic reflections by walls, ceiling and ground are not considered in the present work. As a result, the six input variables describing the problem can be fully represented by four variables, which largely reduce the complexity of the surrogate model. Next, the wave reflections will be taken into account. Due to mutual scattering, the physics behind may become too complicated to approximate the physical system by a single surrogate model. Then, the input space may be divided into subregions, for each of which a surrogate model can be built \cite{friedman1991multivariate,quinlan1992learning}. 

\section*{Acknowledgments}
The first author has been supported in part by the Emergence programme of the Science and Technologies of Information and Communication (STIC) department, University Paris-Saclay.


\end{document}